\documentclass[twocolumn,3p,times,procedia]{elsarticle}
\usepackage{bm}
\usepackage{soul}
\usepackage[numbers]{natbib}
\usepackage{ecrc}
\usepackage{titlesec}
\usepackage{float}
\usepackage{color}
\usepackage{setspace}
\usepackage{ragged2e}
\usepackage{multirow}
\usepackage{subfig}
\usepackage{graphicx}
\usepackage[justification=centering]{caption}
\usepackage[ruled]{algorithm2e}
\usepackage{bm}
\usepackage{caption}
\usepackage{amsfonts}
\usepackage{longtable}

\volume{00}

\firstpage{1}

\runauth{}

\jid{procs}

\CopyrightLine{2015}{Published by Elsevier Ltd.}

\usepackage{amssymb}
\usepackage{amsmath}
\usepackage{multicol}

\usepackage[figuresright]{rotating}

\usepackage{fancyhdr}

\fancyheadoffset[RO,EL]{0pt}
\usepackage{geometry}

\begin{document}

\begin{frontmatter}




\dochead{}
\title{
\begin{flushleft}
{\bf \Huge Diversified and Compatible Web APIs Recommendation in IoT}
\end{flushleft}
}
 %

\author[]{\bf \Large \leftline {Wenwen Gong$^a$, Huiping Wu$^b$, Xiaokang Wang$^c$, Xuyun Zhang$^d$, Yawei Wang$^a$}
		\bf \Large \leftline {Yifei Chen$^*$$^a$, Mohammad R. Khosravi$^{e,f}$}}

\address{\bf  \leftline {$^a$Colleage of Information and Electrical Engineering,
China Agricultural University, Beijing 100000, China}

\bf  \leftline {$^b$Blockchain Laboratory of Agricultural Vegetables,
Weifang University of Science and Technology, Shouguang 262700, China}

\bf  \leftline {$^c$Department of Computer Science, 
St. Francis Xavier University, Antigonish, NS, Canada}

\bf  \leftline {$^d$Department of Computing, 
Macquarie University, Sydney NSW 2122, Australia}

\bf  \leftline {$^e$Department of Computer Engineering, 
Persian Gulf University, Bushehr 7516913817, Iran}

\bf  \leftline {$^f$Department of Electrical and Electronic Engineering, Shiraz University of Technology, Shiraz 71557-13876, Iran}
}

\fntext[]{Wenwen Gong is with Colleage of Information and Electrical Engineering, China Agricultural University, Beijing 100000, China (email: wen.gong@cau.edu.cn).}

\fntext[]{Huiping Wu is with Blockchain Laboratory of Agricultural Vegetables, Weifang University of Science and Technology, Shouguang 262700, China (email: wuhp760731@wfust.edu.cn).}

\fntext[]{Xiaokang Wang is with Department of Computer Science, St. Francis Xavier University, Antigonish B2G, Canada (email: xkwang@stfx.ca).}

\fntext[]{Xuyun Zhang is with Department of Computing, Macquarie University, Sydney NSW 2122, Australia (email: xuyun.zhang@mq.edu.au)}

\fntext[]{Yawei Wang is with Colleage of Information and Electrical Engineering,
	China Agricultural University, Beijing 100000, China (email: yaweiwang@cau.edu.cn).}

\cortext[]{Yifei Chen (Corresponding author) is with Colleage of Information and Electrical Engineering,
China Agricultural University, Beijing 100000, China and  will handle correspondence at all stages of refereeing and publication, also post publication (email:glhfei@cau.edu.cn).}

\fntext[]{Mohammad R. Khosravi is with Department of Computer Engineering, Persian Gulf University, Bushehr 7516913817, Iran and Department of Electrical and Electronic Engineering, Shiraz University of Technology, Shiraz 71557-13876, Iran (email: m.khosravi@mehr.pgu.ac.ir)}
\begin{abstract}
With the ever-increasing popularity of Internet of Things (IoT), massive enterprises are attempting to encapsulate their developed outcomes into various lightweight web APIs (application programming interfaces) that can be accessible remotely. In this situation, finding and composing a list of existing web APIs that can corporately fulfill the software developers' functional needs have become a promising way to develop a successful mobile app, economically and conveniently. However, the big volume of candidate IoT web APIs put additional burden on the app developers' web APIs selection decision-makings, since it is often a challenging task to simultaneously guarantee the diversity and compatibility of the finally selected a set of web APIs. Considering this challenge and latest successful applications of game theory in IoT, a \underline{Div}ersified and \underline{C}ompatible web \underline{A}PIs \underline{R}ecommendation approach, namely \emph{DivCAR}, is put forward in this paper. First of all, to achieve API diversity, \emph{DivCAR} employs random walk sampling technique on a pre-built ``API-API'' correlation graph to generate diverse ``API-API'' correlation subgraphs. Afterwards, with the diverse ``API-API'' correlation subgraphs, we model the compatible web APIs recommendation problem to be a minimum group Steiner tree search problem. Through solving the minimum group Steiner tree search problem, manifold sets of compatible and diverse web APIs ranked are returned to the app developers. At last, we design and enact a set of experiments on a real-world dataset crawled from \emph{www.programmableWeb.com}. Experimental results validate the effectiveness and efficiency of our proposed \emph{DivCAR} approach in balancing the web APIs recommendation diversity and compatibility.
\end{abstract}


\end{frontmatter}
\maketitle

\section{Introduction} 
Internet of Things (IoT) describes the seamless interconnectivity among machines and human devices which gather and share massive information. With the wide adoption of IoT, Service-oriented Architecture
(SoA) and other novel technologies, the latest decade has witnessed the birth of web service sharing platforms, such as online  \emph{www.programmableWeb.com}\footnote{http://www.programmableweb.com}, which hosts a wide variety of lightweight web APIs (Application Programming Interfaces) published by various services vendors \cite{Almarimi2019Web}. Up to September 2020, the largest web APIs ecosystem, \emph{programmableWeb}, gathers at least 23,612 published web APIs belonging to more than 400 predeﬁned categories. All these sharing platforms usually offer the external invocation function of published web APIs to developers \cite{Cao2020Integrated}. For app developers, development cycle and cost can be saved by reusing remotely these third-party web APIs \cite{Hao2017service} and combining a few of them into value-added applications \cite{2009Context}. 

Benefiting from IoT applications in various fields, the continuous evolution of the web API economy allows developers to find desired web APIs and further integrate them into a mashup in IoT by resorting to exact keyword-matching techniques \cite{Jiang2015Exact}. However, the massive candidate web APIs with similar functionality but distinct quality would often make it hard for app developers to select the suitable web APIs, especially for those developers who do not have much background knowledge of web APIs. For example, if a developer intends to accomplish an app with three functions \{\emph{``Video'', ``Blogging'', ``Photos''}\}, he/she will search for a set of collectively-satisfied candidate web APIs over 23,612 web APIs from the \emph{programmableWeb.com} repository through feeding in these three keywords successively. Then, the website would respectively return corresponding 1087, 753 and 661 functional-qualified web APIs, which means that such app would require exhaustive exploration of nearly $ 1000^{3} $ web API compositions. As many researchers have pointed out, finding the optimal one from so many combinations is known as classical NP-hard \cite{Ardagna2006business, Ali2017Search}. In this case, how to recommend top-\emph{K} compositions to app developers remains a non-trivial task. Therefore, it happens that there is a rapid growth in the need for recommendation system (RS) technique \cite{Gong2018} to fulfill multifarious software products including apps.

Although a large body of efforts have been made in current researches in this field, we still identify several deficiencies:

(1) First of all, there is a significant lack of diversity\footnote{There are two kinds of diversity in recommendation system: aggregate or individual diversity. The aggregate diversity is for all users across all recommended items while the individual diversity is for each individual user. Here, our focus is on the aggregate diversity.} \cite{2017KBSDiversityReview} in existing methods since most of them place too much emphasis on accuracy. Moreover, they usually exhibit the redundant web APIs owing to sharing uniform web APIs in web APIs recommendation lists. This repetitive invocations for identical web API easily result in decreasing the rate of customer's satisfaction and increasing cost of a little extra resource to some extent.

(2) In the second place, taking into account the tense time-to-market limit, it's impractical for app developers to inspect the official mannuals of all potentially-possible web APIs and confirm the compatibility between them. Thus, top-\emph{K} combinations with the same functionality but distinct compatibility are different to catch, which is prone to reduce the usefulness of recommendations and the productivity of developers.

Recently, game theory, as an model of strategic interaction among rational decision-makers, has been widely applied to various problems in IoT, such as allocate resources, assign tasks and so on. In view of these two limitations for automatic app development, we put forward a novel \underline{Div}ersity-aware and \underline{C}ompatibility-driven web \underline{A}PIs \underline{R}ecommendation method (called \emph{DivCAR}) in this paper. Our contributions of this paper are chiefly summarized as follows:

(1) We introduce the idea of sampling to achieve the diversity of web APIs recommendation. To the best of our knowledge, this is the first effort to combine the idea of sampling with minimum group Steiner tree search algorithm for the diversity of web APIs recommendation. 

(2) We conduct an effective web APIs recommendation algorithm to return the top-\emph{K} useful composition solutions in terms of comprehensive diversity-accuracy measure.

(3) We performed a series of systematic experiments over a real-world dataset crawled from the website \emph{programmableWeb.com}, and then exported experiment results reveal the superiority of our proposal than comparision methods.

The remainder of our paper is structured as follows. Section 2 investigates and classifies relevant research works. In order to better facilitate an understanding of our approach, a motivating example is described intuitively in Section 3. Key notations and their meanings required by our algorithm are presented in section 4. Our recommendation solution \emph{DivCAR}, in Section 5, is discussed in detail. In Section 6, we verify the effectiveness of our approach through the exported experimental results. Last but not least, Section 7 draws a conclusion and points out future work.

\section{Related Work}

In recent years, a growing number of scholars have devoted themselves to the multi-angle researches on web APIs-based app development from theoretical research to practical application in IoT, which lays a solid groundwork for our solution. In this section, we would summarize existing literatures on web APIs recommendation for app creation from the perspectives of accuracy, diversity and compatibility.

\subsection{Accuracy-oriented Recommendation}

The accuracy in IoT web service ranking or recommendation has been highly concerned by researchers \cite{2007QoS, 2013Recommending, 2015SegevUnified}. In the reusable composition context, Yao et al. propose a probabilistic matrix factorization approach with implicit correlation regularization to explore web API recommendation for mashups. Experimental results over a large-scale real-world service dataset demonstrate that their method outperforms the state-of-the-art collaborative filtering approaches in terms of accuracy. In \cite{Hao2017service}, Hao et al. put forward a method named targeted reconstructing IoT service descriptions (TRSDs) for the sake of more valuable information hidden in mashup description; then, according to that, a novel service recommendation algorithm is developed to advance accuracy by 6.5$\%$. Afterwards, to improve the quality of the recommendation results, Zhong et al. \cite{Zhong2018Web} enhance the above approach by dynamically reconstructing objective service profiles and design a novel recommendation strategy based on similar profiles. A topic-adaptive web APIs recommendation method integrating multi-dimensional information, called hierarchical Dirichlet processes-factorization machines (HDP-FM), is proposed in \cite{Li2018Top-adaptive}, which achieves a good accuracy performance in terms of precision, recall, F-measure and NDCG. Deep learning is introduced in \cite{2018XiongDeeplearning} for web service recommendation by Xiong and Wang et al.; in concrete, the complex invocation interactions are integrated into a deep neural network through combining collaborative filtering and textual content to bring forth improvement at precision rate. Like this work, \cite{2018Dual} also utilizes deep learning and matrix factorization to do an in-depth mining of textual item content. Recently, Huang et.al \cite{huang2021deep} introduce a novel deep reinforcement learning-based approach to deal with the long-term recommendation problem in IoT, which significantly outperforms existing methods in terms of Hit-Rate and NDCG. Besides, \cite{tang2020novel} concerns security in android applications. However, accuracy is often not the only focus of app developers in evaluating the recommended web APIs. Therefore, it is also of practical significance to explore other important recommendation performance criteria besides accuracy.  
\subsection{Diversity-oriented Recommendation}
Diversity, a key mertric for evaluating the recommendation performances in IoT settings, can significantly expand the end users' service selection scope and as a result, enhance the end users' satisfaction degree \cite{2017KBSDiversityReview}. To name a few, through clustering, a category-aware distributed service recommendation (CDSR) model is put forward in \cite{Xia2015Category} based on an app requirements in the form of textual input. This method not only enhances the accuracy to some extent but also gains better diversity in long-tail recommendation. Similarly in \cite{Gao2015Manifold}, following the idea ``services should be recommended cooperatively, not individually'', the authors of \cite{Gao2017A} continue to extend their work by exploiting the variant vKmeans cluster technique based on K-means algorithm and updating service ranking mechanism. Then, a novel framework for service set recommendation (SSR) is proposed to provide more diverse recommendations for developers. Literature \cite{Gu2016Service}  brings forth a service package recommendation (SPR) to produce recommendations with more selective scope by exploiting a similar discourse analysis technique in \cite{Gao2017A}. To generate a list of diverse web APIs, Kang et al. \cite{Kang2016Diversifying} incorporate functional similarity, QoS (Quality of Service) values \cite{jin2019time} and diversity features of web services into a \emph{web services graph} to rank web service diversity degree. Finally, top-\emph{K} web services with sound diversity are returned app developers. However, the above-mentioned approaches can only a set of web services, instead of outputting multiple sets of web services, which often narrows users' web services selection scope. Considering this drawback, Cheng et al. \cite{Cheng2020Diversified} refine the previous work in \cite{Kang2016Diversifying} to find more diverse web services lists. Different from \cite{Kang2016Diversifying} where each node in \emph{web service graph} is an individual web service, the each node in \cite{Cheng2020Diversified} is web service lists. In addition, Gu et al. \cite{2017Diversity} propose a novel diversity-optimization method based on a time-sensitive semantic cover tree (T2SCT) to make diversified recommendations with pretty little compromise on accuracy. Then, to achieve both accurcy and diversity, He et al. \cite{He2020Diversified} use Matrix Factorization (MF) to predict useful third-party web APIs for app development. Recently, a novel method called \emph{DivRec\_LSH} is proposed in \cite{2020KBSWang} to achieve diverse and efficient recommendations through Locality-Sensitive Hashing (LSH) technique. Although the above solutions can achieve diverse recommendations, they often suffer from low compatibility of returned web services, which makes application development less successful.

\subsection{Compatibility-oriented Recommendation}

In IoT web APIs-based app creation, compatibility is often a critical metric to measure that the collaboration performances among web APIs and hence gains ever-increasing attention. Here, we need to point out that plus papers \cite{2018YaoMashup}, \cite{He2020Diversified} and \cite{Cheng2020Diversified}, the next few recent approaches to be explained are all based on the co-invocation data in IoT scenario. And the same is true of our research in this paper. In \cite{Huang2015Model}, the correlation graph describing web APIs is modeled to excavate the compatibility-aware evolution patterns of web APIs in IoT environment. The authors exploit a quantitative method of tag-based semantic matching-degree between inputs and outputs of two web APIs to evaluate the web API compatibility. However, this method as well as the measurement in \cite{Chen2018Goal} are prone to introduce unless edges when modeling the service correlation for such input/output-based compatibility evaluation. Inspired by this issue, Qi et al. \cite{Qi2020Data} have made great efforts to define a new compatibility metric which an ``API-API'' correlation graph is established. Finally, combined with minimum group Steiner tree search approach, a compatibility-aware APIs recommendation method is proposed to return a set of functionally-qualified web APIs with high-level compatibility. Afterwards, in \cite{Qi2019Finding}, an updated weighted APIs correlation graph is put forward through introducing edges' weights (lager is better). Furthermore, the authors of \cite{Qi2019Finding} propose a weight-aware compatibility web APIs recommendation method. Nevertheless, these approaches fail to provide developers with multiple sets of compatible web APIs. To cope with this issue, in \cite{Gong2020web}, Gong et al. continue to make efforts in accomplishing a keywords-driven web APIs group recommendation, which could deliver multiple sets of IoT web APIs which are functional-qualified and compatibility-guaranteed. 

However, work \cite{Gong2020web} still suffers from a low diversity across multiple web APIs recommendation lists as it cannot traverse as more nodes and edges bridging web APIs as possible in the web APIs correlation graph. In view of this drawback, in this paper, we develop a diversity-aware and compatibility-driven web APIs recommendation solution based on the game theory, called \emph{DivCAR}, via exploiting the sampling technique \cite{mahmud2020survey} to carry out more comprehensive coverage of the whole search space. Specific details will be elaborated in Section 5. 

\section{Motivating Example}

\begin{figure*}[htb]
	\centering{\includegraphics[width=5.5 in]{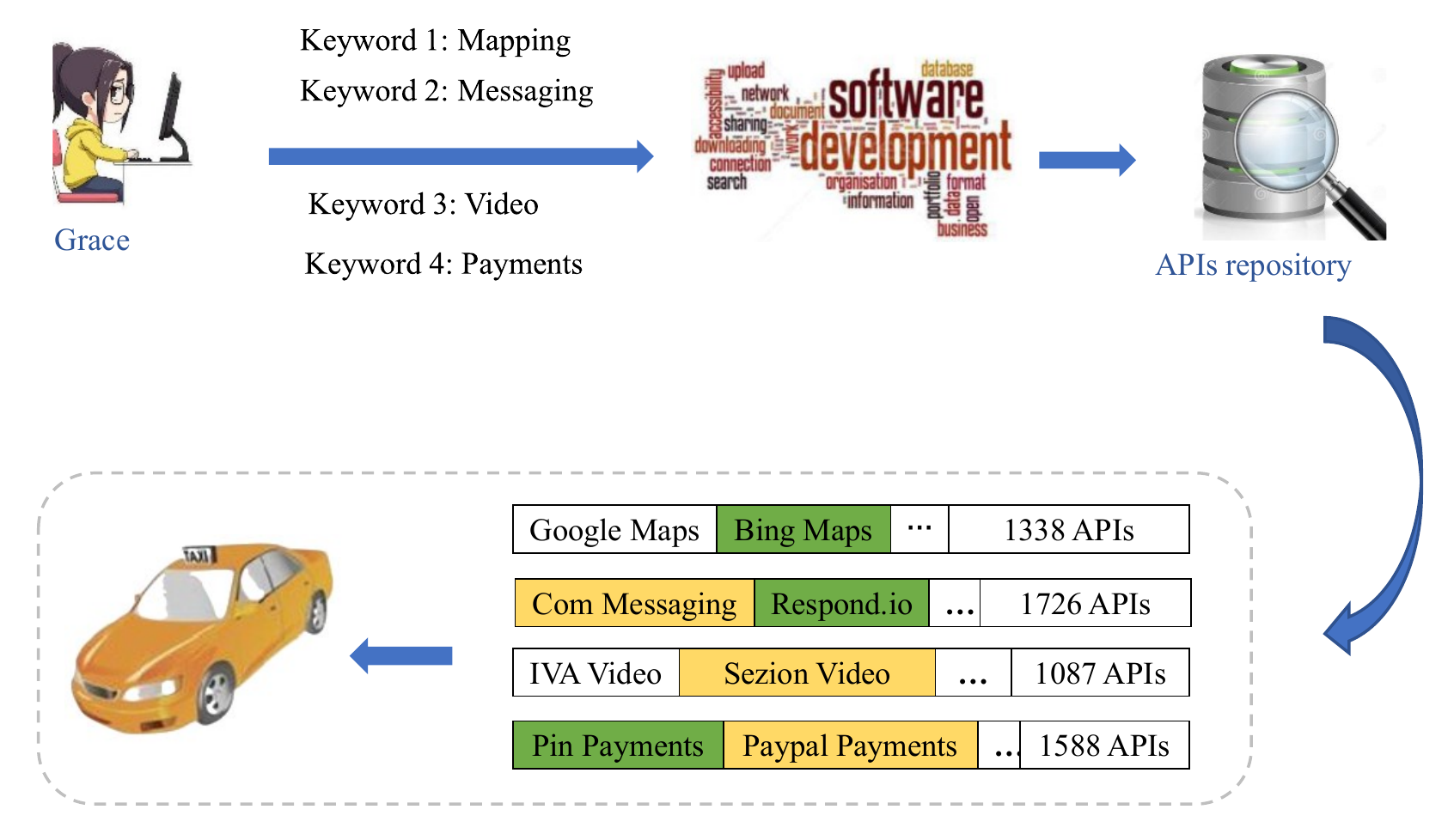}}
	\caption{A motivating example of web APIs recommendation for app creation in IoT.\label{fig_1}}	
\end{figure*}

In this section, we present a real-world IoT example in Figure \ref{fig_1} to clarify the motivation of our research. As shown in Figure \ref{fig_1}, a developer, i.e., \emph{Grace}, intends to develop a mobile taxi-hailing app \cite{khazbak2020preserving} that can aid passengers to quickly find an available taxi. Generally, this app often involves four sub-functions: mapping, messaging, navigating and payment. Thus, when developing such a mobile taxi-hailing app, \emph{Grace} needs to enter a set of keywords, \{\emph{``Mapping'', ``Messaging'', ``Navigating'', ``Payments''}\} into web APIs search engines (\emph{ProgrammableWeb.com}) to search for a set of qualified APIs that can collectively satisfy \emph{Grace}'s functional requirements. Here, Figure \ref{fig_1} exhibits the candidate web APIs for each keyword as well as their possible combinations  (marked by various colors).

However, in the above keywords-driven mobile app development scenario, two challenges are often raised. First of all, \emph{Grace} may know little about the compatibility among the returned web APIs by the APIs recommender system, while less-compatible web APIs may lead to a high failure rate when composing these APIs into a complex app. Therefore, from the perspective of \emph{Grace}, it is becoming a necessity to guarantee that the returned a set of web APIs are not only functional-qualified but also compatible enough. Second, to provide \emph{Grace} more flexibilities when choosing appropriate web APIs, it is significant for the recommender system to return multiple sets of qualified APIs, instead of only one set of APIs. This way, \emph{Grace} can pick out her preferred API set from multiple candidate sets, so as to reduce the APP development cost and accelerate the development speed.

As a result, in this situation, how to recommend manifold diversity-aware web APIs lists with functionality and compatibility guarantee is becoming a challenging and meaningful issue that deserves intensive research.

\section{Problem Definition}

In the following section, we summarize the rationale of necessary terms for the process of web APIs recommendation solutions in IoT settings. Key notations and their meanings are presented in Table 1.

\textbf{Definition 1 (Web APIs Ecosystem):}
A IoT web APIs ecosystem saves plentiful web ``APP-API'' interaction information. Here, a IoT web APIs ecosystem, is defined as $ S = (API, A) $ where $ API = \left\lbrace api_1, api_2, ..., api_n \right\rbrace  $ and $ A = \left\lbrace a_1, a_2, ..., a_m \right\rbrace $ denote the collection of web APIs and apps in the IoT web APIs ecosystem (i.e., \emph{ProgrammableWeb.com}) separately. An ``APP-API'' co-usage record $ s_i \in S $ exists on condition that some web API $ api_i \in API $ is successfully invoked in an app $ a_i \in A $ by an app developer.

\textbf{Definition 2 (Keywords Query):}
Given a IoT web APIs ecosystem $ S $, keywords query is denoted as $ Q = \left\lbrace q_1, q_2, ..., q_r \right\rbrace $ from category attribute of web APIs in $ S $, which represents the end-users' functional requirements for expected apps.

\textbf{Definition 3 (Vertices):}
A collection of vertices is represented by $ V = \left\lbrace v_1, v_2, ..., v_n \right\rbrace $ corresponding to a set of web APIs $ API $ in web APIs ecosystem $ S $, in which each vertex covers a group of keywords query $ \left\lbrace q_1, q_2, ..., q_r \right\rbrace $.

\textbf{Definition 4 (Edges):}
Given a cluster of vertices $ V = \left\lbrace v_1, v_2, ..., v_n \right\rbrace $, there are a set of corresponding edges defined as $ E = \left\lbrace e_1, e_2, ..., e_s \right\rbrace (s \leq n)$. If a pair of $ v_i $ and $ v_j $ have ever been appeared in an identical app, an edge $ e(v_i, v_j) $ is added into set $ E $. That is to say, suppose each app $ a_i \in A $ invokes all web APIs in $ API $ and a complete graph associated with the cluster of vertices $ V = \left\lbrace v_1, v_2, ..., v_n\right\rbrace $ would be acquired.

\begin{figure}[htb]
	\centering{\includegraphics[width=3.3 in]{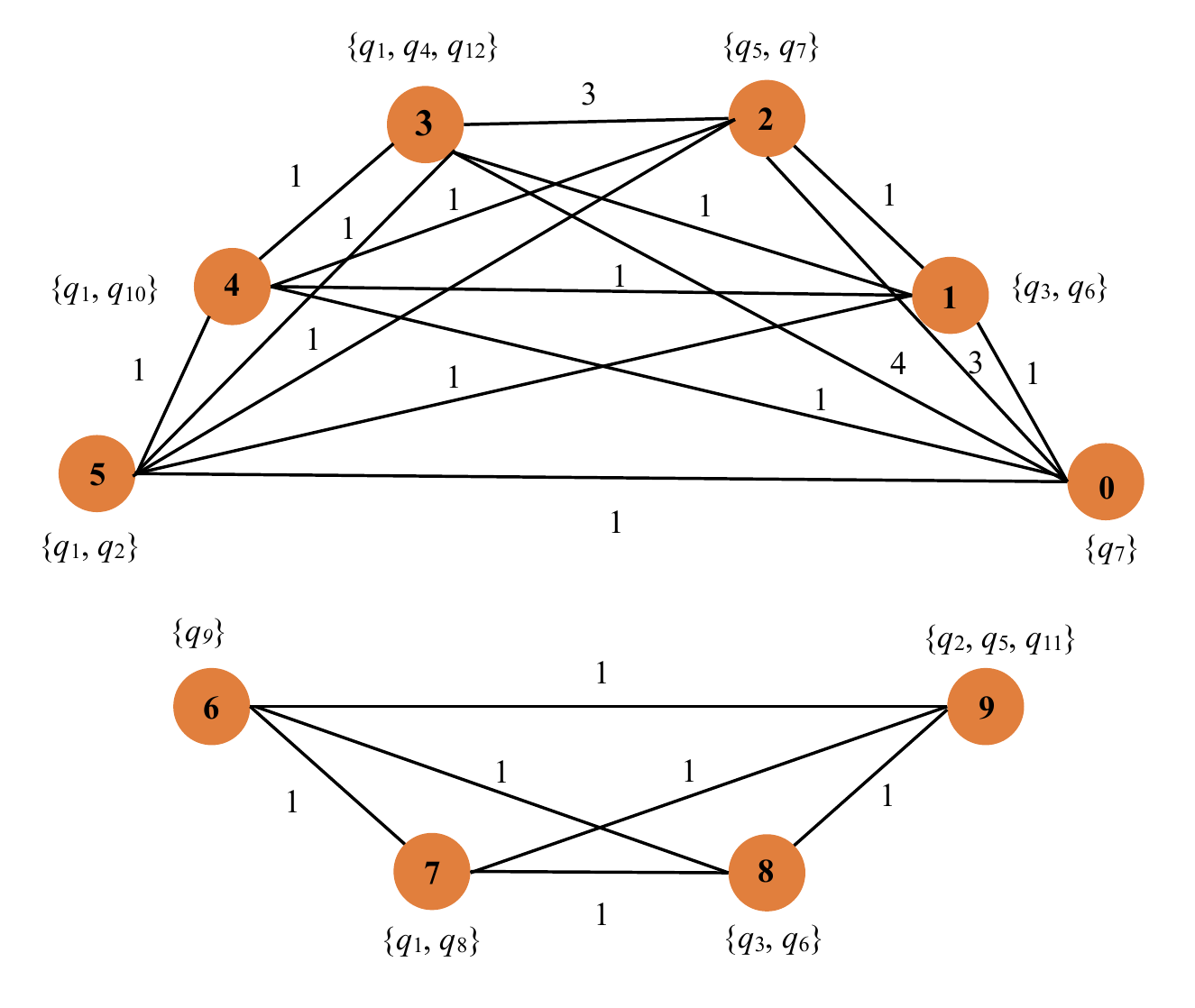}}
	\caption{The partial weighted web APIs correlation graph.\label{fig_2}}	
\end{figure}

\textbf{Definition 5 (Compatibility):}
In this paper, one of the facets that interest us is the number of times that a pair of $ v_i $ and $ v_j $ have ever been integrated successfully into identical app according to historical app development, which is called as the compatibility value $ c_{i, j} $ (an integer greater than zero) of an edge $ e(v_i, v_j) $. In some ways, the value of compatibility for an edge reflects the weight or popularity between the two web APIs allied to an edge. As depicted in Figure \ref{fig_2}, $ c_{2, 3} \textgreater c_{3, 4} $, then we can draw the conclusion that $ v_3 $ has better compatibility with $ v_2 $ instead of $ v_4 $. At this level, the granularity of web APIs versions and so on is out of the scope of this article.

\textbf{Definition 6 (Diversity):}
Assume that two sets of web APIs collection $ API_1, API_2 $, there is a diversity value $ D = 1- \frac{\left| API_1 \cap API_2 \right|}{\left| API_1 \right| + \left| API_2 \right|} $ indicating the degree to which the web APIs from $ API_1, API_2 $ are not similar to some extent. For example, considering two web APIs collections $ API_1 = {api_1, api_2, api_3}, API_2 = {api_1, api_4} $, the diversity value equals to 0.8.

\textbf{Definition 7 (Weighted Web APIs Correlation Graph (W-ACG)):}
Each app published on the \emph{ProgrammableWeb} website indicates a meaningful integration for constituted web APIs. Above definitions and such valuable information allow a web APIs correlation graph W-ACG = \emph{G(V, E, W)} quoted from \cite{Qi2019Finding} to be established offline. Specifically, we employ the example in Figure \ref{fig_2} to explain for easing readers' understanding. In this example, there are a total of 10 vertices where each vertex $ A_i (\in A)(0 \le i \le 9)$ (marked in dark orange) represents a web API covering a collection of functional keywords, e.g., $ A_3 $ possesses the keywords describing functions $ \left\lbrace q_1, q_4, q_{12} \right\rbrace $ while $ A_0 $ can fulfill the function set $ \left\lbrace q_7 \right\rbrace $. There exists an edge between these two vertices with a compatibility of 0.25 that is taken reciprocal, which says the weight between is 4 and they have ever been integrated collectively four times. Note here that the ``$ A_i $'' in Figure \ref{fig_2} is shortened to ``i'' for brevity's sake.

In addition, as you can see in Figure \ref{fig_2}, there are two unconnected subgraphs due to a fraction of IoT web APIs from different domains, e.g. \emph{print service} and \emph{health}. In the real world, it would be almost impossible for an app developer to enter such irrelevant keywords that belong to different domains. Therefore, the maximal connected subgraph serves as our W-ACG.

As a result, our recommendation problem can be formalized as follows: given a set of keywords query instances $ Q $ in the web APIs ecosystem $ S $, it requires us to find a set of potential web APIs compositions with minimum compatibility $ C $ while guaranteeing accuracy and diversity constraints $ P $, which can be described as in (1):

\begin{equation}
\begin{split}
& \mathit{Maximize\ C\ satisfying\ P} \\ 
& C = \left\lbrace C_1, ..., C_K\right\rbrace, C_l = C_{api_1} + ...+ C_{api_n}, l \in \left\lbrace 1, ..., K \right\rbrace  \\ 
& P = \left\lbrace P_1, ..., P_K\right\rbrace, P_1 \ge ... \ge P_K
\end{split}	
\end{equation}

Here, $ API_i, API_j \in \left\lbrace API_1, ..., API_K\right\rbrace $ means two arbitrary recommended web APIs lists from web APIs compositions $ \left\lbrace API_1, ..., API_K \right\rbrace $. Constraints $ P $ means putting accuracy first. That is, diversity is considered when the accuracy of the list is the same since diversity is ensured during the sampling process in \emph{Step 1}.

\begin{table}[!htb]
	\renewcommand{\arraystretch}{1.3}
	\caption{Specification of symbols used in this paper}
	\label{table1}
	\centering
	\begin{tabular}{l|l}
		\hline
		\bfseries Symbol & \bfseries Specification \\
		\hline
		$ S = (API, A) $ & A web APIs ecosystem \\
		\hline
		$ API $ & The web APIs collection\\
		\hline
		$ A $ & The collection of apps \\
		\hline
		$ Q $ & A sequence of required keywords \\
		\hline
		$ V $ & The vertex collection  \\
		\hline
		$ E $ & The undirected edge collection  \\
		\hline
		$ W $ & The weight collection \\
		\hline
		$ G(V, E, W) $ & Weighted web APIs correlation graph \\
		\hline
		$ G_{sam} $ & A cluster of subgraphs $ \left\lbrace G_1, ..., G_z \right\rbrace $ \\
		\hline
		$ p $ & The size of each sample \\
		\hline
		$ \mathit{Q' \subseteq Q} $ & A state of \emph{Q} in our model\\
		\hline
		\emph{$ N_{key} $} & A pre-built keyword nodes set \\
		\hline
		\emph{N\_Set} & The neighboring nodes set of $v_i$ \\
		\hline
		$ N_{sam} $  & The collection of sampled vertexes \\
		\hline
		\multirow{2}*{\emph{R}} & The priority queue recording transitive \\
		& trees \\
		\hline
		\multirow{2}*{\emph{RT}} & The priority queue recording candidate \\
		& trees \\
		\hline
		\multirow{2}*{\emph{ST}} & The list of optimal trees from each \\
		& subgraph \\
		\hline
		$T(v_i, Q')$ & A status in our algorithm\\
		\hline
		$T_{min}(v_i, Q)$ & Minimum group Steiner tree \\
		\hline
		$T_{div}$ & Diverse top-\emph{K} Steiner trees \\
		\hline
	\end{tabular}
\end{table}

\begin{algorithm}[h]
	\caption{\emph{DivFinder-Sampling (G, Q)}}
	\LinesNumbered 
	\KwIn{\\
		\emph{G(A, E, W): weighted APIs correlation graph}; \\ 
		$ Q = \left\lbrace q_1, …, q_r \right\rbrace $ \emph{: a set of query keywords} \\}
	\KwOut{\\
		$ G_{sam} = \left\lbrace G_1, \cdots, G_z\right\rbrace $ \emph{: a cluster of weighted correlation subgraphs } \\}
	
	$ N_{key}  = generate\_nodes(G, Q) $ \\
	\For{each $ i \in z $}{
		$ N_{sam} = \emptyset $ \\ 
		\emph{randomly select a node $ v_i $ from $ N_{key}$}\\
		\emph{add $ v_i $ into $ N_{sam} $}\\
		\emph{update $ N_{sam} $}\\
		\While{$ \left| N_{sam} \right| \textless \left| G_i \right| $}{
			\emph{N\_Set = $ \emptyset $} \\
			\If {$ w_{v_i, v_j} \textgreater 0 $}{
				\emph{add $ v_j $ into N\_Set}\\
				\emph{update N\_Set } \\
			}	
			\emph{randomly select a neighbor $ N_i $ from N\_Set into $ N_{sam} $ } \\
			\emph{add $ N_i $ into $ N_{sam} $ } \\
			\emph{update $ N_{sam} $ } \\
		}	
		
		\emph{build new subgraph $ G_i $ using $ N_{sam} $} \\		
	}
	\emph{return $ G_{sam} = \left\lbrace G_1, \cdots, G_z\right\rbrace $ }
	
\end{algorithm}

\section{Our Recommendation Solution: \emph{DivCAR}}

Our recommendation solution \emph{DivCAR}, in this section, is discussed in detail. Before we get into the details, let's first describe the Steiner tree in subsection 5.1, a key technique used in the solution. And then, the specific steps of our recommendation approach \emph{DivCAR} are depicted in subsection 5.2.

\subsection{Steiner tree}
The Steiner tree (ST) or minimum Steiner tree, named after Jakob Steiner, refers to a combinatorial optimization solution and has been proven to be computationally hard and NP-complete \cite{Garey1977rectilinear, Hwang1992The}. The Steiner tree problem is superficially similar to the minimum spanning tree problem, in that both of them interconnect a given set of vertexes by a graph of shortest length, where the length is the sum of the lengths of all edges. It has been proven that the shortest interconnect is exactly a tree. However, the difference between them is that extra intermediate vertexes and edges could be added into the graph in the Steiner tree problem in order to reduce the length of the spanning tree. In other words, the minimum spanning tree can be recognized as a special case of the minimum Steiner tree. 

Let us provide a mathematical definition of the Steiner tree. Given \emph{G(V, E, W)} and $ V'\subseteq V $, then $ T(v_i, Q') $ is called a Steiner tree of $ V' $ in $ G $ iff both conditions in (1) and (2) hold: (1) keywords of one vertex in $ V' $ are different from those of another one from $ V' $; (2) $ Q' $ contains required keywords in set $ Q $, i.e., satisfying $ Q' \subseteq Q $. In the application scenario of this paper, however, there is usually more than one corresponding node for each required keyword from $ Q $, and thus the original Steiner tree problem no longer meets our needs so that the group Steiner tree needs to be introduced. The essential difference between them is that the same keywords of distinct vertexes in $ V' $ exist. We will consider the example in Figure \ref{fig_2}, supposing that $ Q = \left\lbrace q_1, q_2, q_3, q_5 \right\rbrace $, and then $ T(v_2, Q')$ is a Steiner tree of $ V' $ where $ Q' = \left\lbrace q_1, q_2, q_3, q_5 \right\rbrace $, $ V' = \left\lbrace v_5, v_1, v_2 \right\rbrace $ and edges are $ \left\lbrace e(v_5, v_1), e(v_1, v_2)\right\rbrace  $; furthermore, supposing that $ Q = \left\lbrace q_1, q_3, q_7 \right\rbrace $, then there are two group Steiner trees that meet the requirements, i.e., $ T(v_2, Q')$ where $ Q' = \left\lbrace q_1, q_3, q_7 \right\rbrace $, $ V' = \left\lbrace v_5, v_1, v_2 \right\rbrace $ and edges are $ \left\lbrace e(v_5, v_1), e(v_1, v_2)\right\rbrace  $, and $ T(v_0, Q')$ where $ Q' = \left\lbrace q_1, q_3, q_7 \right\rbrace $, $ V' = \left\lbrace v_5, v_1, v_0 \right\rbrace $ and edges are $ \left\lbrace e(v_5, v_1), e(v_1, v_0)\right\rbrace  $. In general, since there are diverse group Steiner trees for a set of initially-given query keywords, this allows us to searh and decide several minimum group Steiner trees, which results in more intricate decision-makings \cite{bhardwaj2021advanced}. In the following subsection, we explain step by step how to address this problem.
\subsection{Diversity-aware and Compatibility-driven Web APIs Recommendation Approach: \emph{DivCAR}}

\begin{figure}[htb]
	\centerline{\includegraphics[width=3.2 in]{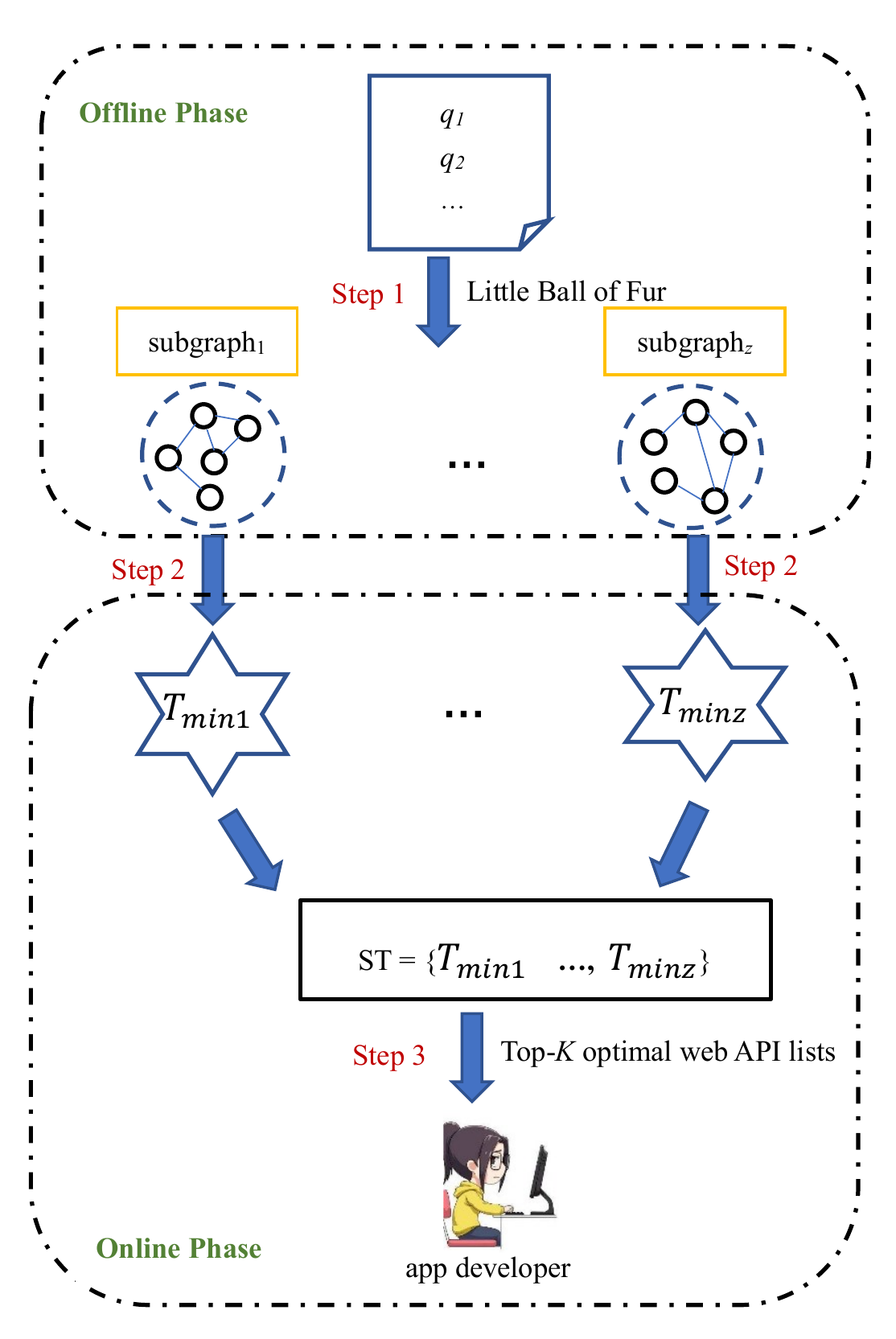}}
	\caption{Working procedure of \emph{DivCAR}.\label{fig_3}}	
\end{figure}

Our solution aims at excavating deeply historical ``APP-API'' invoking information from large web APIs ecosystems $ S $ to reflect ``the wisdom from the crowd''. Specifically, from the data-driven perspective, a wealth of ``APP-API'' invoking data stored in the web server of \emph{ProgrammableWeb.com} carries abundant invocation information between apps and their constituent web APIs, which provides solid background knowledge for professionally building apps. To further achieve diversity across all web APIs recommendation compositions, our proposed \emph{DivCAR} is chiefly threefold, as presented in Figure \ref{fig_3}. First, we exploit random walk sampling technique on a pre-established weighted ``API-API'' correlation graph to generate diverse subgraphs. Second, with these various subgraphs, we model the compatible web APIs recommendation process as a minimum group Steiner tree search problem on the basis of graph theory. Finally, through working out the search algorithm, multiple groups of compatible and diverse web APIs are ranked and returned to the app developers.

\vspace{3 pt}
\noindent \textbf{\emph{Step 1: Generate diverse subgraphs of W-ACG by random walk sampling}}
\vspace{3 pt}

Our current methods \cite{Gong2020web, Qi2020Data, Qi2019Finding} for information retrieval based on graphs are limited to the most frequently-used web APIs, which hinders the diversity of recommended results. In this step, to further speed up diverse web APIs compositions, sampling strategy for graphs, as an effective technique thoroughly researched in \cite{Leskovec2006Sampling}, is introduced to generate optimal coverage of W-ACG. Furthermore, to produce better representations for W-ACG, \cite{Leskovec2006Sampling} answers two natural questions: (a) Which sampling techniques can perform well with various existing sampling methods; (b) How many the sample size can be. Overall, sampling strategy based on simple random node selection could perform well. Morover, this approach can also work well as the sample size of the evolutionary subgraph pattern is reduced to approximately 15 percent of the original graph, which will be verified in Section 6.

Concretely, we employ an open-source Python library that consists of more than twenty algorithms for graph sampling, \emph{Little Ball of Fur} \cite{Rozemberczki2020Little}, to attain multiple subgraphs. In a nutshell, it can be called a Swiss Army knife for sampling tasks from graph structured data. In the first place, \emph{Little Ball of Fur} is developed through various techniques for node, edge, and exploration-based graph sampling. In the second place, it provides a unified application public interface which makes the application of sampling algorithms trivial for end-users. The most attractive advantage of \emph{Little Ball of Fur} is that it can leave the vertex index values unchanged.

In consideration of the characteristics of our dataset, the \emph{Random Walk Sampler} among them is allowed to serve for our algorithm. Moreover, we all know that this often results in a great number of sampling times if we expect to gain representative and desired results, due to the inherent uncertainty and probability of sampling technique. Without loss of generality, in the experimentation of this paper, we sample each set of keyword query sequences $ Q = \left\lbrace q_1, q_2, ..., q_r \right\rbrace $ 100 times by \emph{Random Walk Sampler} of \emph{Little Ball of Fur} based on the prebuilt W-ACG, which is represented by $ \left\lbrace G_1, \cdots, G_{100}\right\rbrace $. We will analyze the influence of sampling times $ z $ on the experimental results in detail in Section \textcolor{red}{6}. To ensure better understanding, we use \textbf{Algorithm 1} to elaborate the details. 

In addition, since each web API from the \emph{programmableWeb} sharing platform often possesses multiple tags that represent its functions, we collect these tags of all web APIs integrated in each app to form distinct sets of query keywords. What needs to be emphasized here is that all of these groups of query keywords are established in advance. In the meantime, all corresponding subgraphs can be built offline to ensure execution efficiency. Once these subgraphs have been built offline, they will remain relatively stable and can still be updated with minimal overhead.

\vspace{3 pt}
\noindent \textbf{\emph{Step 2: Compatibility-aware web APIs recommendation online}}
\vspace{3 pt}

In the second step, according to end-users' required keywords online, our algorithm will return a compatibility-optimal web APIs composition based on each of 100 subgraphs obtained by \textbf{\emph{Step 1}}. For instance, $ T(v_0, Q')$ and $ T(v_2, Q')$ in Figure \ref{fig_2} mentioned previously. That is to say, given a query $ Q $, this step of \emph{DivCAR} returns a compatibility-optimal minimum group Steiner tree, $ T_{min}(v_i, Q) $, rooted at $ v_i $, while covering each requirement in $ Q $. To be specific, we will state the details below. Here, please note that it is necessary to convert the compatibility value by taking the inverse of $ c_{i, j} $ of edge $ e(v_i, v_j) $ since our object is to try our best to obtain the  ``minimum value'' case of compatibility-aware optimization problem as formulate in equation (1).

\begin{figure}[htb]
	\centerline{\includegraphics[width=3 in]{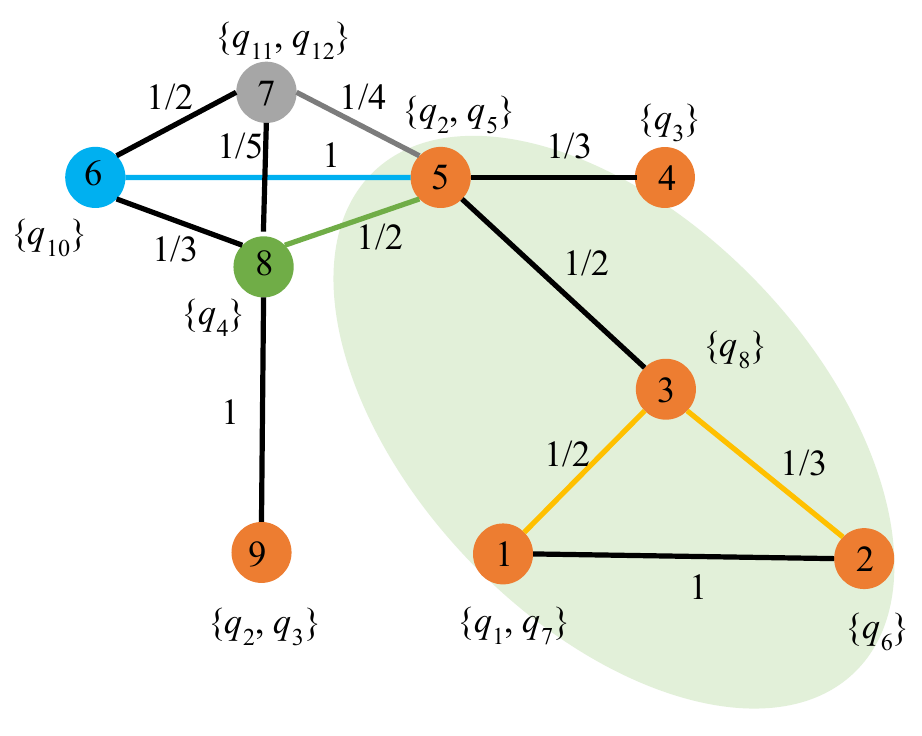}}
	\caption{The \emph{tree growth} and \emph{tree merging} processes: an example.\label{fig_4}}	
\end{figure}
\begin{algorithm}[!h]
	\caption{\emph{DivFinder-DivSearching ($ G_{sam} $, Q)}}
	\LinesNumbered 
	\KwIn{\\
		$ G_{sam} = \left\lbrace G_1, \cdots, G_z\right\rbrace $ \emph{: a cluster of weighted web APIs correlation subgraphs}; \\ 
		$ Q = \left\lbrace q_1, …, q_r\right\rbrace $ \emph{: a set of query keywords}}
	\KwOut{\\
		$ ST = {T_{min_1}, \cdots, T_{min_z}} $: \emph{a list of minimum group Steiner trees answered from all of subgraphs.} \\}
	
	$ R = \emptyset, RT = \emptyset $ \\
	\For{each $ G_i \in G_{sam} $}{
		\For{each $ v_i \in V $}{
			$ Q' = Q \cap C_{v_i} $ \\ 
			\If{$ Q' \ne \emptyset $}{
				\emph{enqueue newtree $T(v_i, Q') $ into R}
			}
		}
		\While{$ R \ne \emptyset $}{
			\emph{dequeue R as $ T(v_i, Q') $} \\
			\If{$ Q' = Q $}{
				\emph{enqueue $ T(v_i, Q') $ into RT} \\
				continue
			}	
			\text{\% the growth operation for one tree } \\
			\For{each $ u \in U(v_i)$}{
				\If{$ w_{u, v_i} + w_{T(u, Q')} \textless w_{T(v_i, Q')} $}{
					$ T(v_i, Q') = e(u, v_i) + T(u, Q') $
					\emph{enqueue $T(v_i, Q')$ into R}
				}
			}
			\text{\% the merging operation for one tree} \\
			\For{each $ T(v_i, Q_1'), T(v_i, Q_2') $}{
				\If{$ Q_1' \cap Q_2' = \emptyset $}{
					\If{$ w_{T(v_i, Q_1')} + w_{T(v_i, Q_2')} \textless  w_{T(v_i, Q_1' \cup Q_2')} $}{$ T(v_i, Q_1' \cup Q_2') = T(v_i, Q_1') \oplus T(v_i, Q_2') $ \\
						\emph{enqueue $ T(v_i, Q_1' \cup Q_2') $ into R}
					}
					
				}
			}
		}
		$ T_{min_i} = RT.top() $ \\
		\emph{add $ T_{min_i} $ into ST} \\
	}
	\emph{return ST}
\end{algorithm}
\begin{algorithm}[!h]
	\caption{\emph{DivFinder-Ranking (ST)}}
	\LinesNumbered 
	\KwIn{\\
		\emph{ST: The list of optimal trees derived from all subgraphs}; \\ 
	}
	\KwOut{\\
		$ T_{div} = \left\lbrace T_{div_1}, \cdots, T_{div_k} \right\rbrace $ \emph{: final recommendation resulting trees} \\}
	
	\emph{R\_ST  = ranking(ST)} \\
	\For{each $ R\_ST_i, R\_ST_j \in R\_ST $}{		
		\If {$ diversity_{R\_ST_i, R\_ST_j} \le \theta $}{
			\emph{add $ R\_ST_i, R\_ST_j $ into} $ T_{div} $ \\
			\emph{update $ T_{div} $}}	
	}   		
	\emph{return} $ T_{div} $	
\end{algorithm}
As depicted in \textbf{Algorithm 2}, according to the theory of the Steiner tree, there are two crucial operations: \emph{tree growth} and \emph{tree merging}, which are described by lines 15-19 and 21-28 of \textbf{Algorithm 2}, respectively. Let us illustrate them with the example in Figure \ref{fig_4}. Concretely, assume that an app developer is entitled to enter a set of required keywords $ Q = \left\lbrace q_1, q_3, q_5, q_6 \right\rbrace $. Let the descending priority queues $ R, RT $ save possible trees and final trees, respectively.

\begin{figure*}[htb]
	\centering
	\subfloat[Tree growth]{\includegraphics[width=3 in]{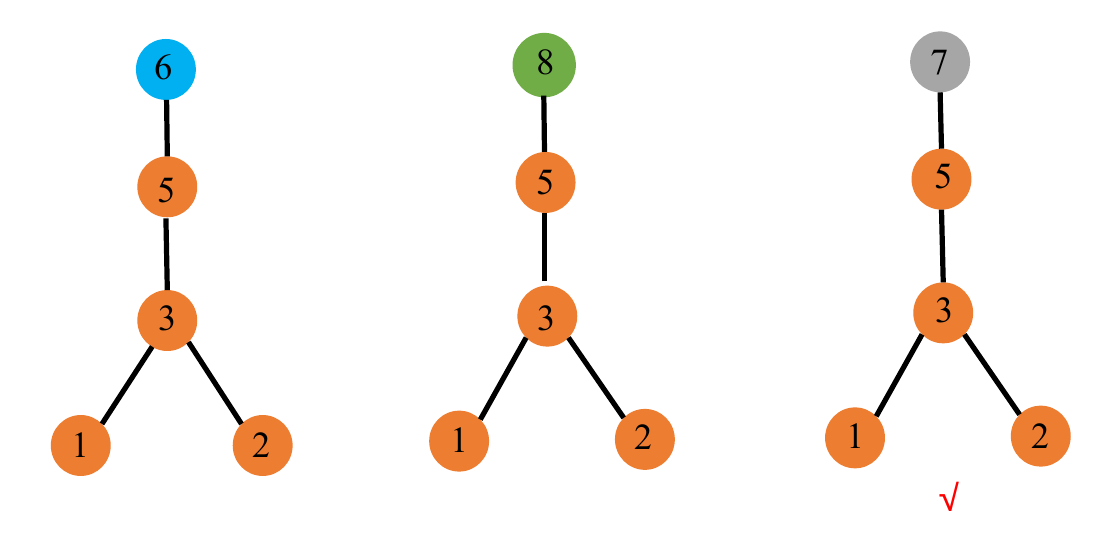}}
	\label{growth}
	\subfloat[Tree merging]{\includegraphics[width=3 in]{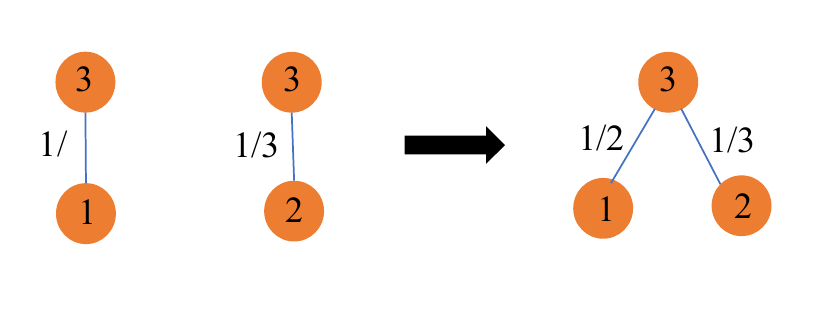}}
	\label{merging}
	\caption{An example for tree operations.\label{fig_5}}
\end{figure*}

As shown in Figure \ref{fig_4}, a transition tree $ T(v_5, Q')$ where $ V' = \left\lbrace v_1, v_2, v_3, v_5 \right\rbrace  $ and $ Q' = \left\lbrace \bm{q_1, q_5, q_6} \right\rbrace $ (keywords are shown in bold) appears at some point in our model (highlighted with light green-shaded area). Afterwards, the algorithm continues further from three directions (marked with different colors in Figure \ref{fig_4}): (1) grows to $ v_6 $ to produce a new tree $ T(v_6, Q')$; (2) extends $ v_7 $ to produce $ T(v_7, Q')$; (3) increases $ v_8 $ to produce $ T(v_8, Q')$. As these three trees are produced with the same keywords $ Q' = \left\lbrace \bm{q_1, q_5, q_6}\right\rbrace $ but distinct compatibility, $ T(v_7, Q')$ with optimal compatibility is enqueued into queue \emph{R} for subsequent operations. Here, please note that although the required query keywords have not been increased, this operation is still a necessary part of the whole algorithm. In this case, $ v_6, v_7$ and $ v_8 $ are referred to as \emph{linking nodes}, while $ v_1, v_2, v_4, v_5 $ and $ v_9 $ are known as \emph{keyword nodes}. The above process is called the \emph{tree growth} operation, which is illustrated in Figure \ref{fig_5} (a) and formalized as Formula (2).

In general, the \emph{tree growth} operation in Figure \ref{fig_5} (a) needs to alternate with the following process called the \emph{tree merging} operation in Figure \ref{fig_5} (b), which is formalized in Formula (3). Among the trees after growth, there often is at least one pair satisfying two conditions: (1) both of them possess identical root nodes; (2) they cover distinct query keywords. In this case, we can attain a better tree with more keywords through merging them to speed up the search process. Likewise, taking Figure \ref{fig_4} as an example, $ T(v_3, Q_1') $ where $ V' = \left\lbrace v_1, v_3 \right\rbrace, Q_1' = \left\lbrace \bm{q_1}, q_7\right\rbrace $ and $ T(v_3, Q_2') $ where $ V' = \left\lbrace v_2, v_3\right\rbrace, Q_2' = \left\lbrace \bm{q_6} \right\rbrace $ (marked in yellow-red) also possess identical root node $ v_3 $, but different keywords. Under the circumstances, we can merge them into a better tree $ T(v_3, Q') $, where $  V' = \left\lbrace v_1, v_2, v_3 \right\rbrace $ and $ Q' = \left\lbrace \bm{q_1, q_6} \right\rbrace $.

Such \emph{tree growth} and \emph{tree merging} operations proceed alternately until an optimal group Steiner tree $ T(v_9, Q) $ where $ V' = \left\lbrace v_1, v_2, v_3, v_5, v_7, v_8, v_9 \right\rbrace$ is finally returned. In addition, it is highlighted that if there is more than one tree with the same keywords root at an identical vertex, then only one tree with the best compatibility will be left.
\begin{center}
	\begin{equation}
	\mathit{w_{T_g(A_i, K')}}= \min_{\substack{\mathit{u \in U(A_i)}}}(\mathit{w_{T(u, K')}} + w_{u,A_i})
	\end{equation}
	\begin{equation}
	\begin{split}
	\mathit{w_{T_m(v_i, Q')}}= \min_{\substack{\mathit{Q = Q_1^{'} \cup Q_2^{'}}, \\ \mathit{Q_1^{'} \cap Q_2^{'} = \emptyset}}}(\mathit{w_{T(v_i, Q_1^{'})}}  + \mathit{w_{T(v_i, Q_2^{'})}})
	\end{split}
	\end{equation}
\end{center}	
\vspace{3 pt}
\noindent \textbf{\emph{Step 3: Diversity and accuracy-aware ranking algorithm}}
\vspace{3 pt}

By means of the previous \textbf{\emph{Step 2}} and \textbf{\emph{Step 3}}, we can acquire 100 top-level group Steiner trees based on all sampled subgraphs, and these trees represent 100 different web APIs compositions. However, theoretically, there must be some compositions with low accuracy due to the contingency inherent in the sampling process, which requires us to rank all of them to pick out high-quality compositions. \textbf{Algorithm 3} presents the pseudo-code of the process of ranking. Specifically, they are first ranked in order of accuracy. Then, the top-\emph{K} web APIs compositions where the diversity between each pair is greater than $ \theta $ are finally returned to app developers as the final web APIs recommendation results.
\section{Experiments}
\subsection{Experimental Configurations}
In our experiments, we employ the real-world dataset crawled from \emph{www.programmableWeb.com} \cite{Qi2020Data}, which includes co-invocation information between 6,146 apps and 18,478 web APIs. According to the redundant co-usage data of the maximum connected graph, W-ACG is prebuilt in advance. Through in-depth mining on the information, we conclude that the vast majority of apps (approximately 96\%) have 2 to 5 keywords. However, what needs special explanation here is that in the research scenario of this article, the apps that only contain two keywords, as a special case, will not be experimented upon. Moreover, all tags for web APIs included in apps are collected as all sorts of keyword sets to generate various subgraphs by the \emph{Little Ball of Fur} library. All experiments are performed on a laptop with identical hardware settings (Intel i5-7300 2.60 GHz CPU, 8.0 GB RAM) and software configurations (Windows 10 and Python 3.7).
\subsection{Performance Metrics}
In this part, we comprehensively measure the performance of \emph{DivCAR} in terms of the following a few widely-utilized metrics:

\textbf{(1) Mean Inter-List Diversity (MILD).}
Inter-list diversity evaluates how different every two lists are on the basis of their Hamming distance (HMD). A higher HMD implies higher diversity between each two lists. It can be calculated as follows:
\begin{equation}
HMD = 1 - \frac{C(i,j)}{\left| RL_i \right|+ \left| RL_j \right| }, i,j \in K \wedge i \ne j
\end{equation}
in which $ \left| RL_i \right|, \left| RL_j \right| $ represent the number of web APIs in $ i_{th}, j_{th} $ recommendation lists, respectively. Besides, if the $ i_{th}, j_{th} $ recommendation lists share no common web API at all, $ C(i,j) =0 $ holds and their \emph{HMD} is 1; otherwise, their \emph{HMD} is 0. 

MILD averages the inter-list diversity across all the recommendation lists in one experiment instance, which is calculated by the formular (5): 
\begin{equation}
MILD = \frac{1}{K(K-1)}\sum HMD
\end{equation}
in which $ K $ is the total number of recommendation compositions.

\textbf{(2) Inner-List Compatibility (MILC).}
Inner-list compatibility measures the success rate of each recommendation compositions, which is the inverse of the value that can be returned directly by our Steiner tree algorithm. And then, MILC averages the inter-list compatibility across all the recommendation compositions in one experiment instance. The larger the value is, the higher the recommendation performance will be.

\textbf{(3) Mean Precision (MP).} Precision of a recommendation list $ RL_i $ is the ratio of correct web APIs in total recommended list. MP is the average of precision across all the recommendation compositions in one experimental instance. As formalized in formula (6), the larger the value is, the better the recommendation accuracy is. 
\begin{equation}
MP = \frac{1}{K}\sum_{i=1}^{K} \frac{\left| RL_i \right| \cap \left| RL_{app} \right| }{\left| RL_i\right| }
\end{equation}
where $ \left| RL_{app} \right| $ denotes the amount of real-world web APIs used in some app.

\textbf{(4) Mean Recall (MR).} Different from precision, recall of a recommendation list is calculated by the ratio of correct web APIs in all real-world web APIs in an app. Similar to MP, larger is better. The calculation of MR is shown in formula (7):
\begin{equation}
MR = \frac{1}{K}\sum_{i=1}^{K} \frac{\left| RL_i \right| \cap \left| RL_{app} \right| }{\left| RL_{app} \right| } 
\end{equation}

The above metrics are calculated for each experiment instance, all within the range of [0, 1]; furthermore, each pair of recommendation lists in each of 100 experimental instances is averaged to measure the performance of our algorithm.

\textbf{(6) Time cost.}
Time cost, as a key metric, is widely-used for the recommendation efficiency; then, the lower the value is, the higher the recommendation efficiency is.

\subsection{Comparative Approaches}
In our experiments, we compare \emph{DivCAR} with the following three representative approaches that all use ``APP-API'' historical co-invocation records:

(1) \emph{SSR} \cite{Gao2017A}: a solution that employs clustering and text analysis techniques to find the \emph{N} sets of web APIs of \emph{N} separate categories that are most similar to a developer's functional input text descriptions. According to popularity, similarity and correlation degree of distinct web APIs, the web APIs compositions with the highest score are selected in the ranking task without considering their diversity and compatibility. Thus, this method is the baseline method in our experiments.

(2) \emph{KC\_MulAGR} \cite{Gong2020web}: a keywords-driven web APIs group recommendation for automatic app service creation, which combines the Steiner tree algorithm without sampling with pairwise ranking based on diversity to return diverse adequate sets of web APIs.

(3) \emph{ATD\_JSC} \cite{Cheng2020Diversified}: is an algorithm that first enumerates all potential web APIs compositions through graph searching. Then, it derives the maximal independent sets (MISs) of similarity graph built from potential solutions to achieve top-\emph{K} diverse web APIs compositions.

(4)\emph{MSD} \cite{MSD2017}: Like in \emph{ATD\_JSC}, \emph{MSD} also first enumerates all possible web APIs compositions through graph searching. Then, it produces top-\emph{K} diverse web APIs compositions through solving max-sum diversification problem.

To perform a more fairer comparison, the optimal parameters in the above four competitive approaches are tuned. Specially, we set diversity threshold $ \alpha = 0.5 $ for ATD-JSC and trade-off parameter $ \lambda =0.5 $ for \emph{MSD}. In our experiments, we conduct the four scenarios where inputted number of keywords $ r $ are 3, 4, 5 and 6, respectively. As for each scenario, we change the values of the parameter sampling times $ z $ from 10 to 100 in steps of 10, which determines the number of sampled subgraphs. And we also vary the parameter sampling size $ p $, which indicates how many the number of vertexes in each subgraph are sampled by \emph{DivCAR}.

\begin{figure*}[htb]
	\centering
	\subfloat{\includegraphics[width=6 in]{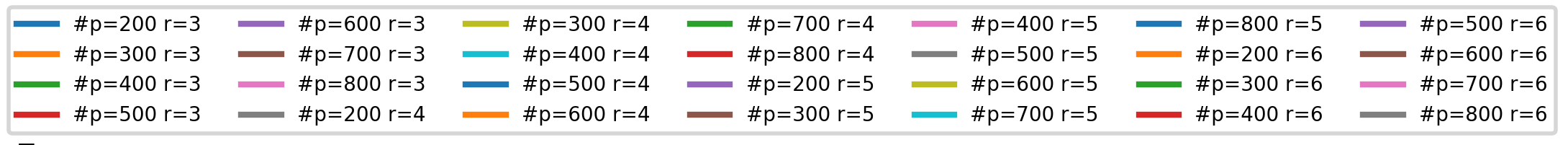}}
	\label{fig9_legend}
	\subfloat[MP convergence]{\includegraphics[width=3 in]{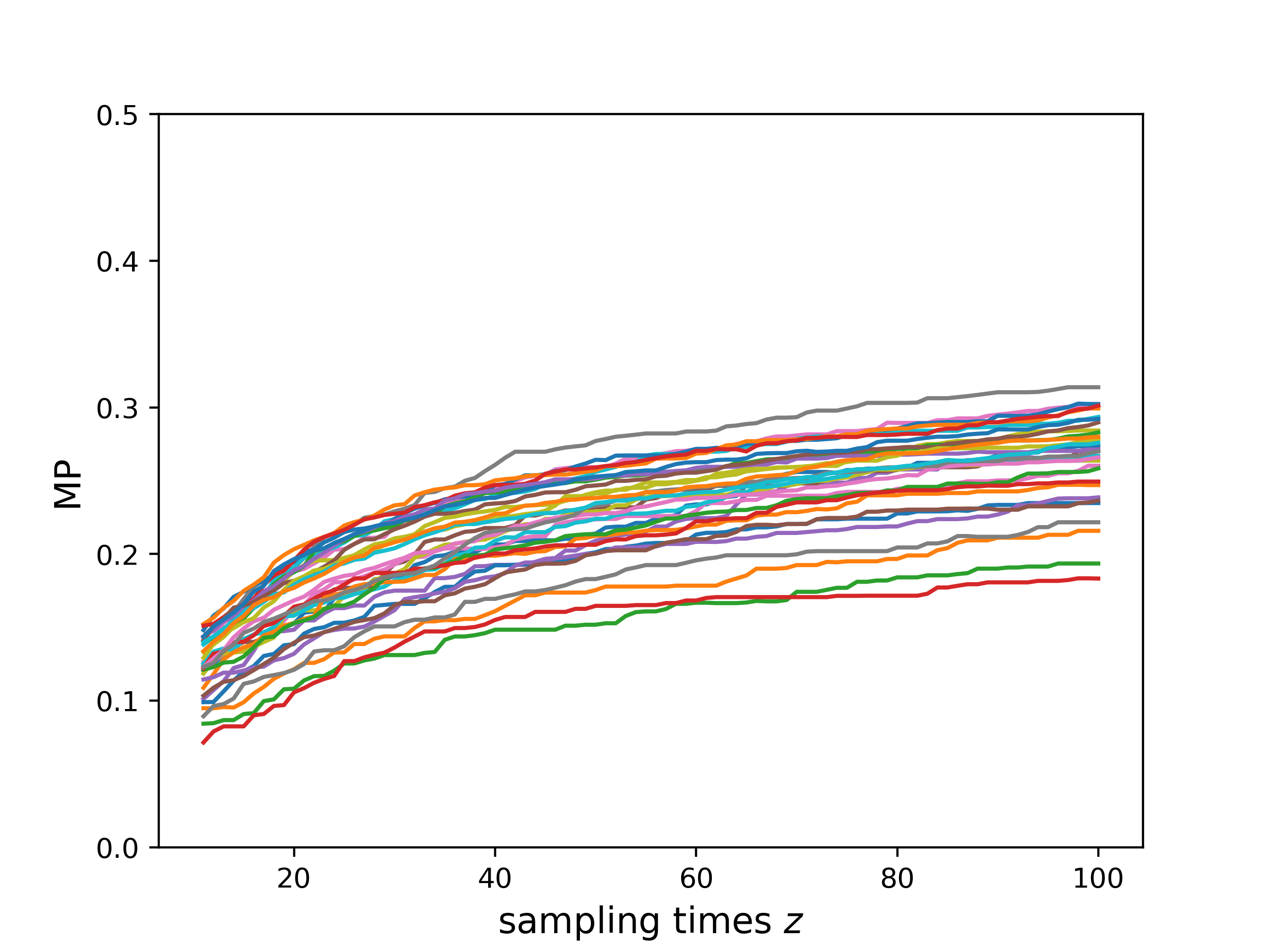}}
	\label{convergence_precison}
	\subfloat[MILD convergence]{\includegraphics[width=3 in]{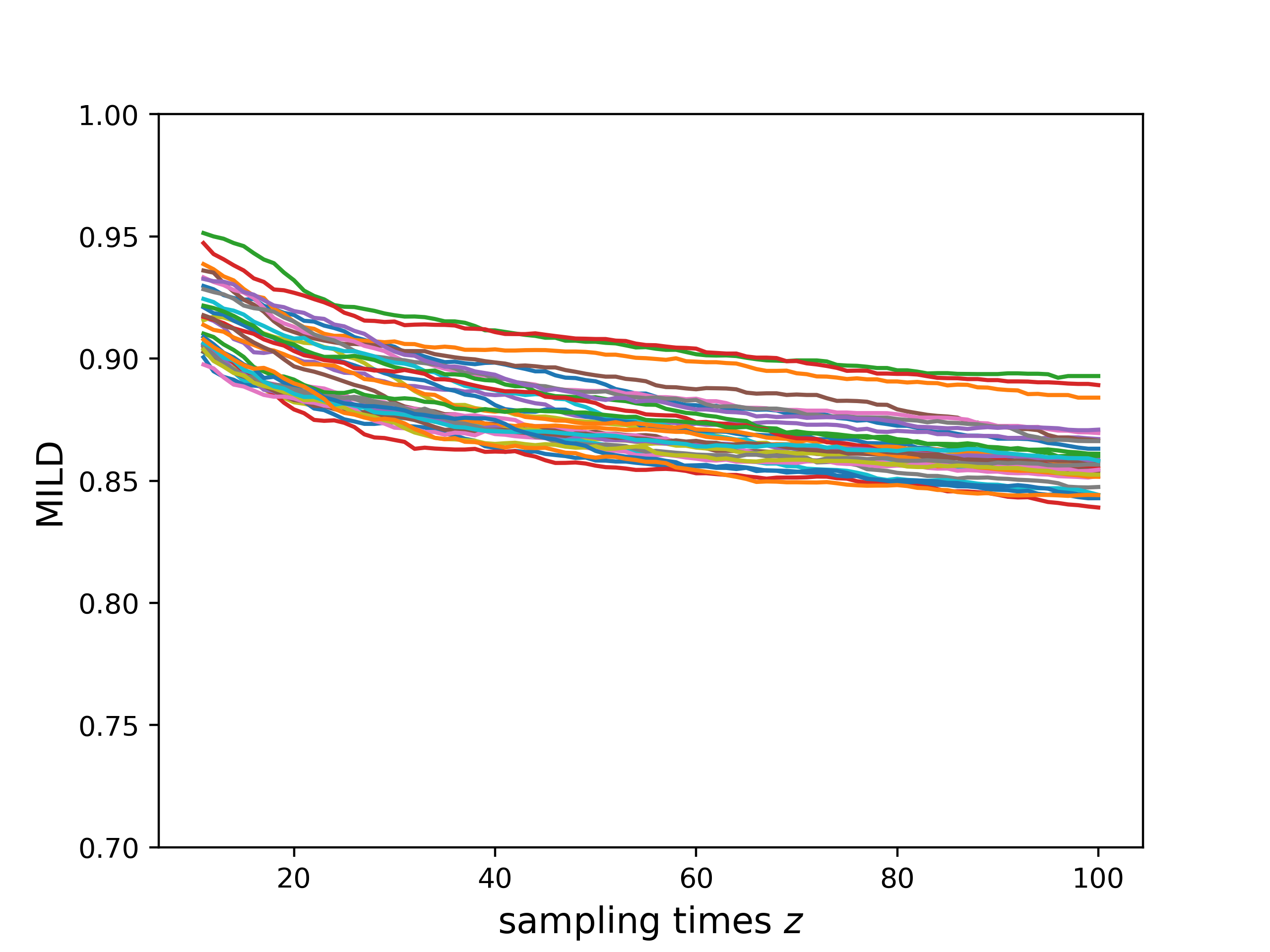}}
	\label{convergence__diversity}
	\caption{Performance convergence of \emph{DivCAR} w.r.t. sampling times.\label{figconvergence}}
\end{figure*}
\begin{figure}[htb]
	\centering{\includegraphics[width=3 in]{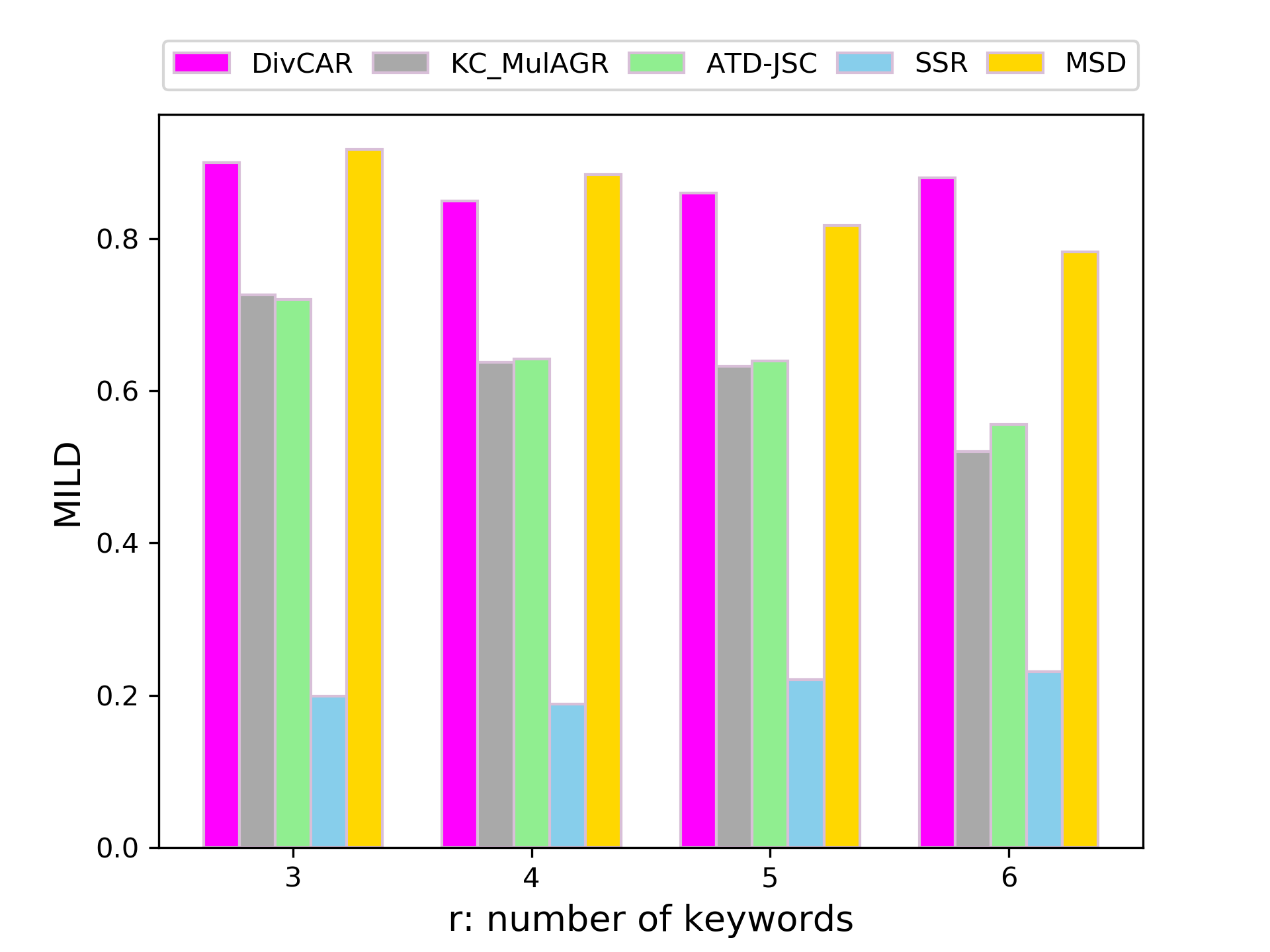}}
	\caption{Recommendation diversity comparison.\label{profile-MILD}}	
\end{figure}
\begin{figure}[htb]
	\centering{\includegraphics[width=3 in]{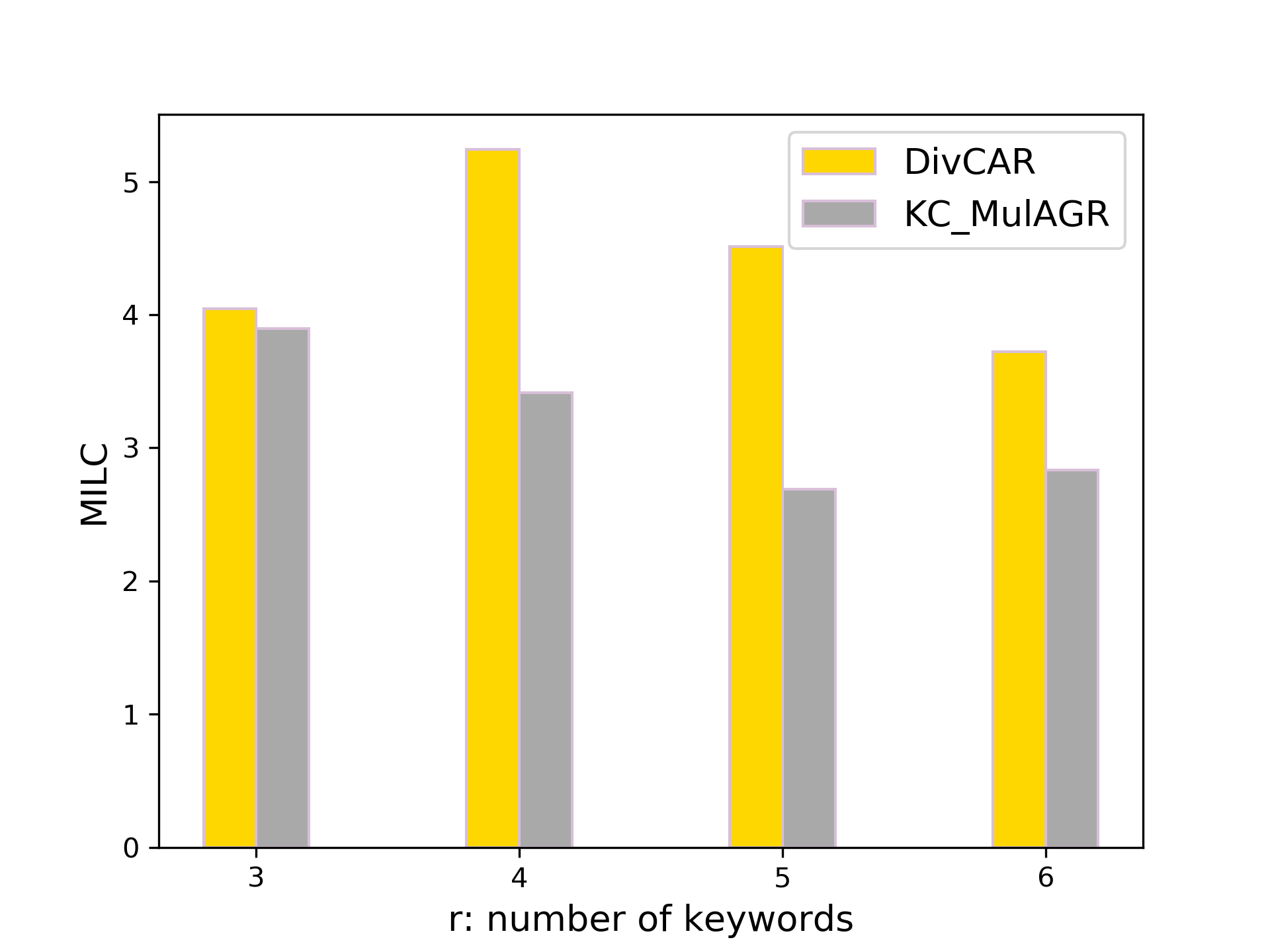}}
	\caption{Recommendation compatibility comparison.\label{profile-MILC}}	
\end{figure}
\begin{figure*}[htb]
	\centering
	\subfloat[MP]{\includegraphics[width=3 in]{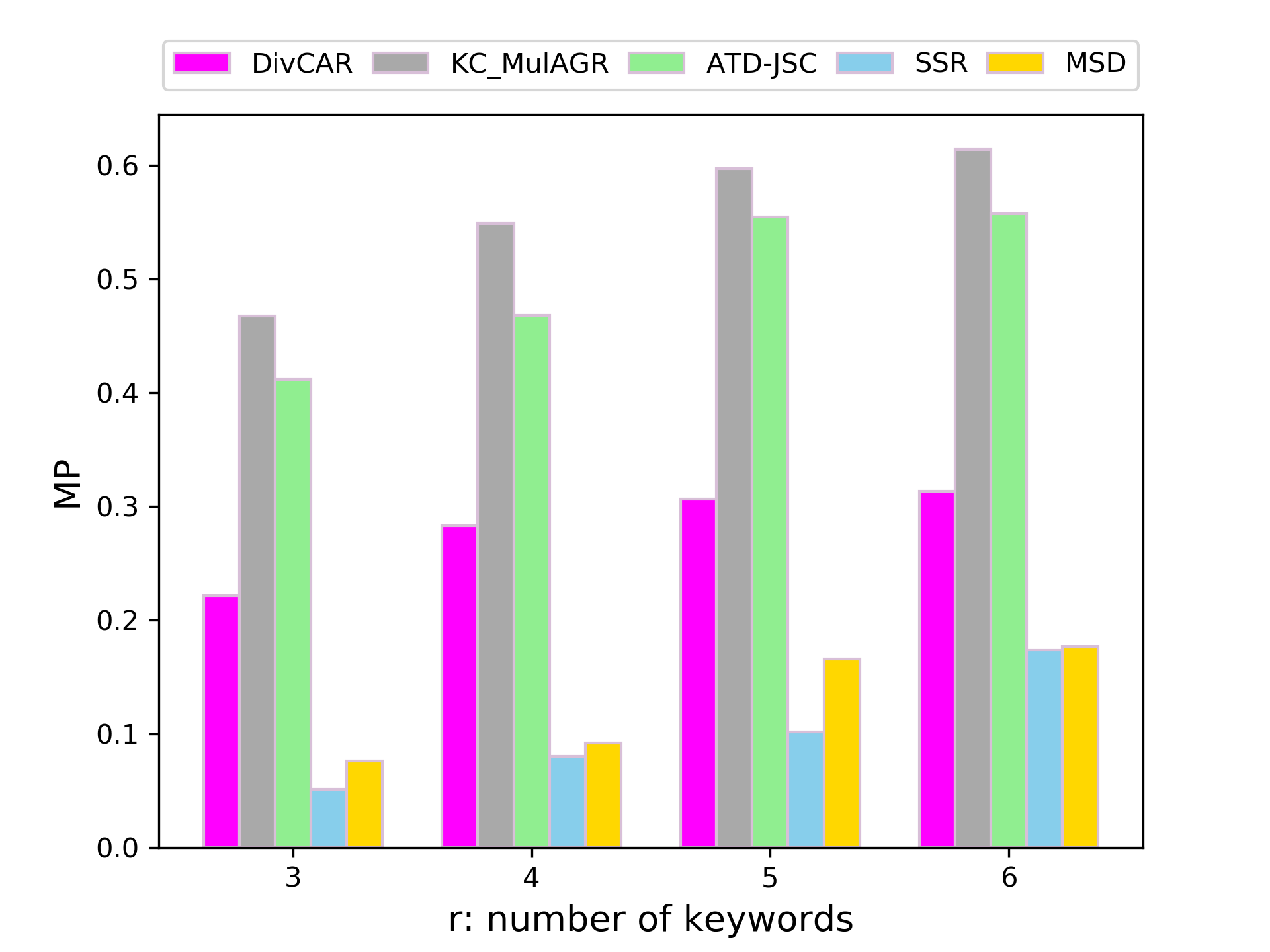}}
	\label{MP}
	\subfloat[MILD]{\includegraphics[width=3 in]{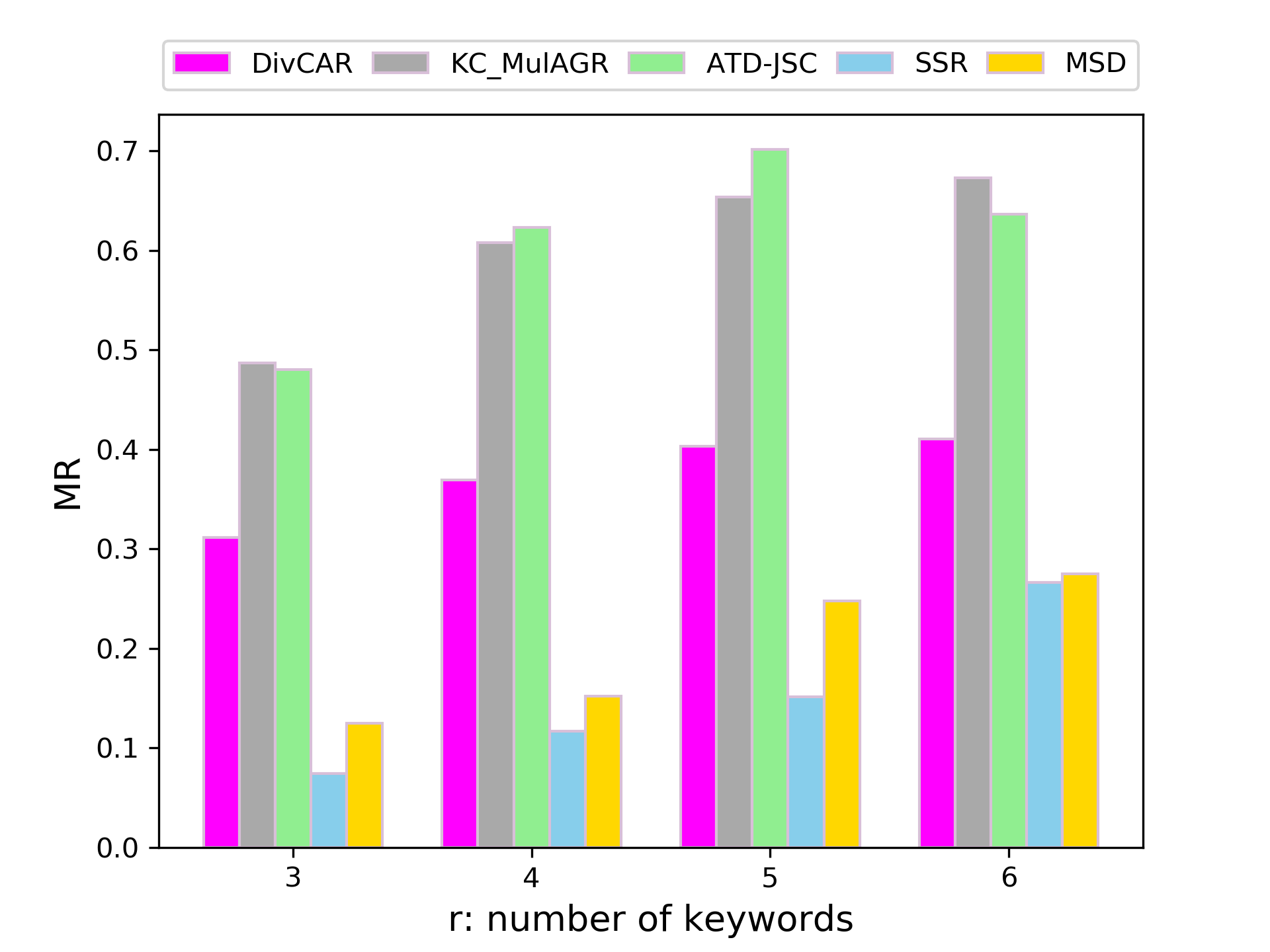}}
	\label{MD}
	\caption{Recommendation accuracy comparison.\label{profile_precision_recall}}
\end{figure*}
\begin{figure}[htb]
	\centering{\includegraphics[width=3 in]{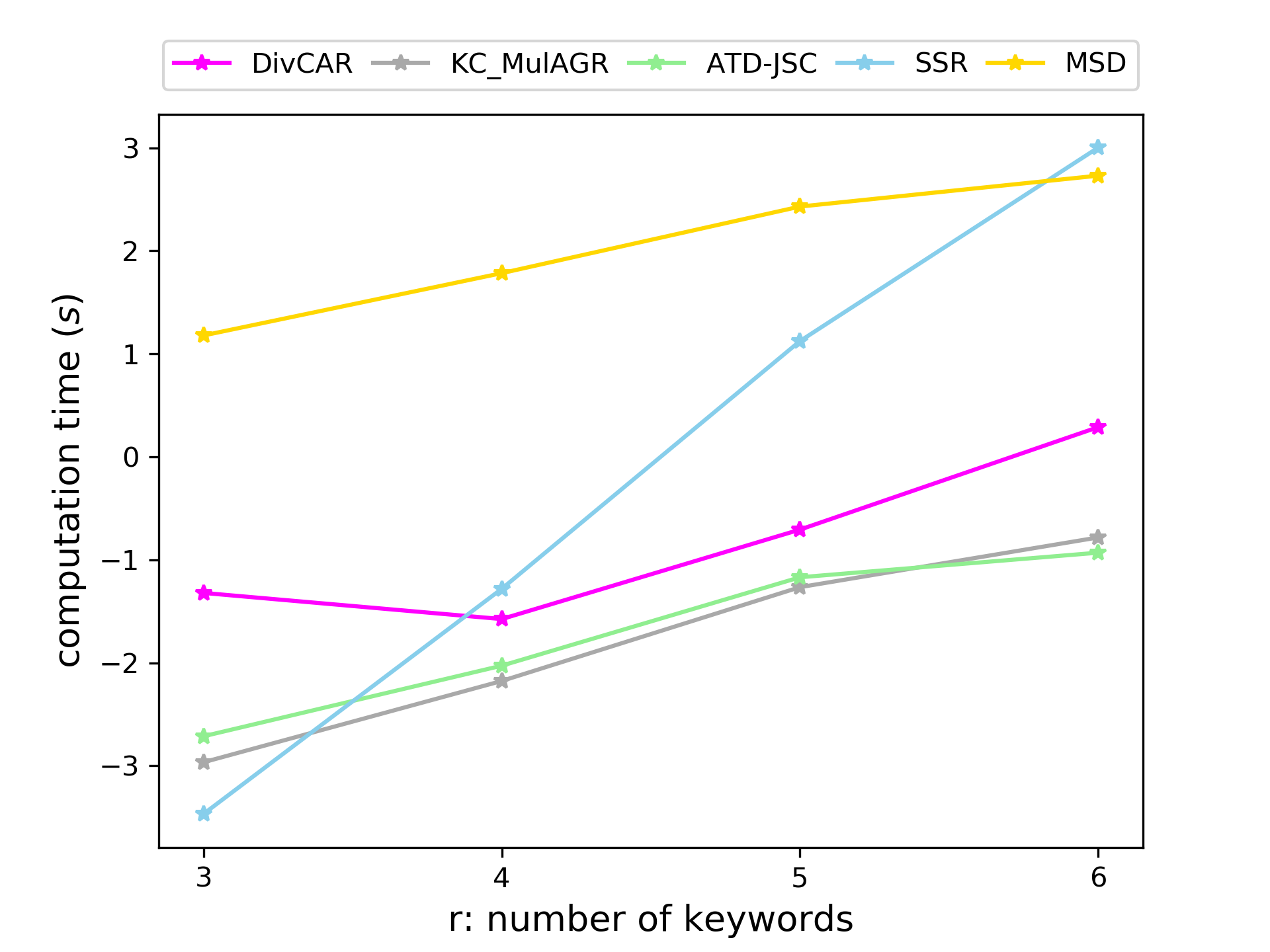}}
	\caption{Computation time comparison.\label{profile-timecost}}	
\end{figure}
\begin{figure*}[htb]
	\centering
	\subfloat[MP convergence]{\centering \includegraphics[width=3 in]{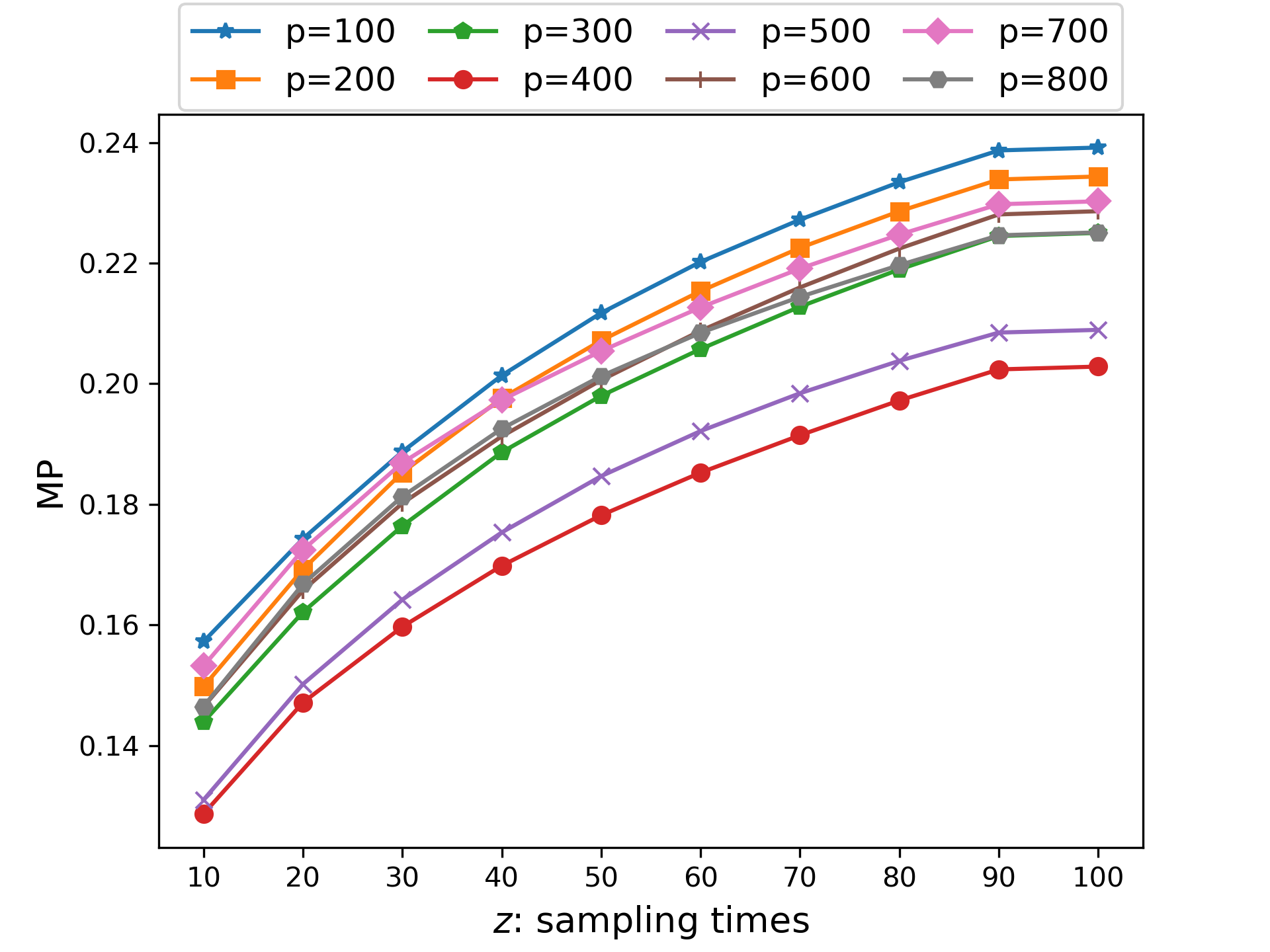}}
	\label{relationMP}
	\subfloat[MILD convergence]{\centering \includegraphics[width=3 in]{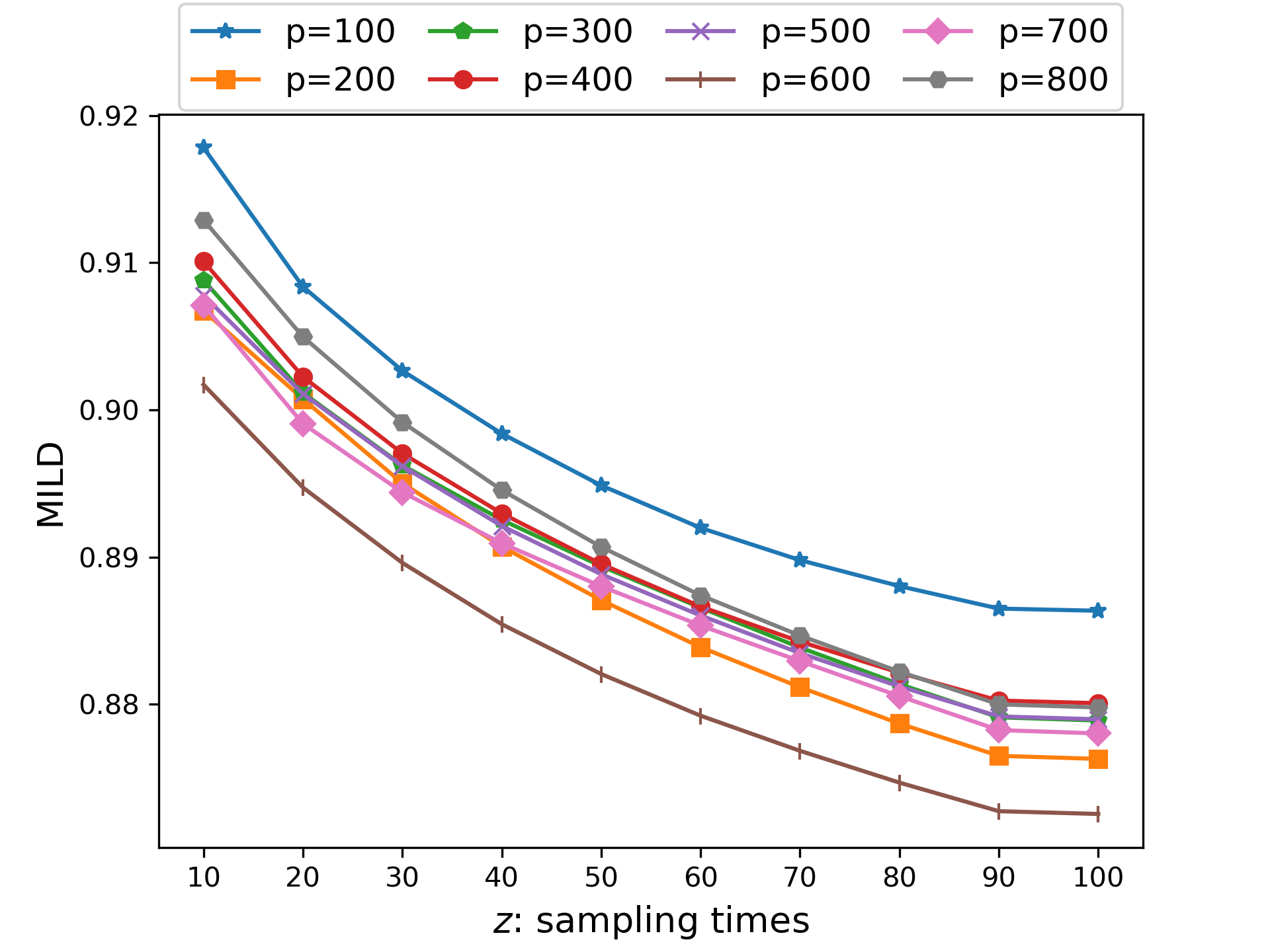}}
	\label{relationMILD}
	
	\subfloat[MP convergence]{\centering \includegraphics[width=3 in]{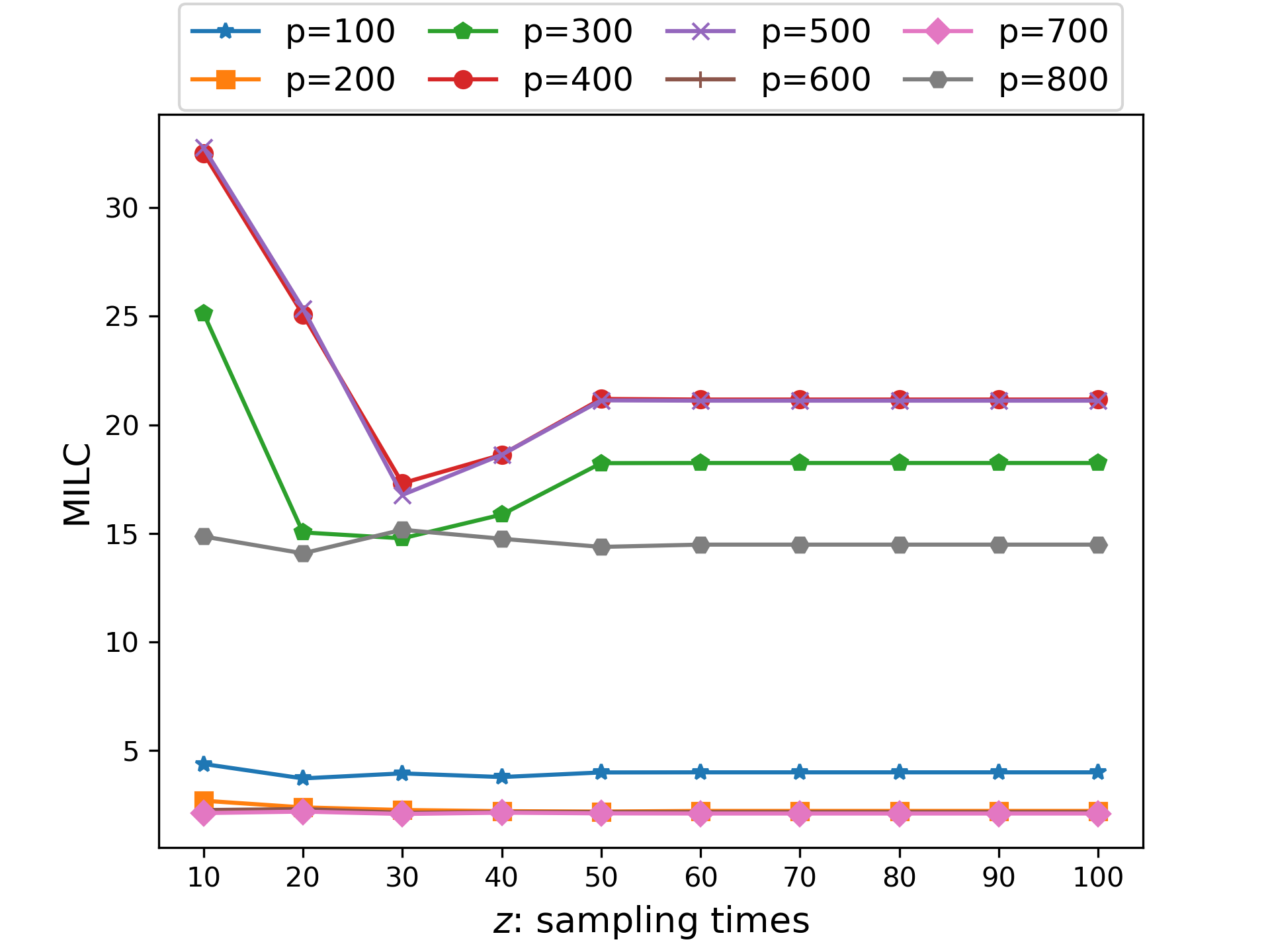}}
	\label{relationMILC}
	\subfloat[MILD convergence]{\centering \includegraphics[width=3 in]{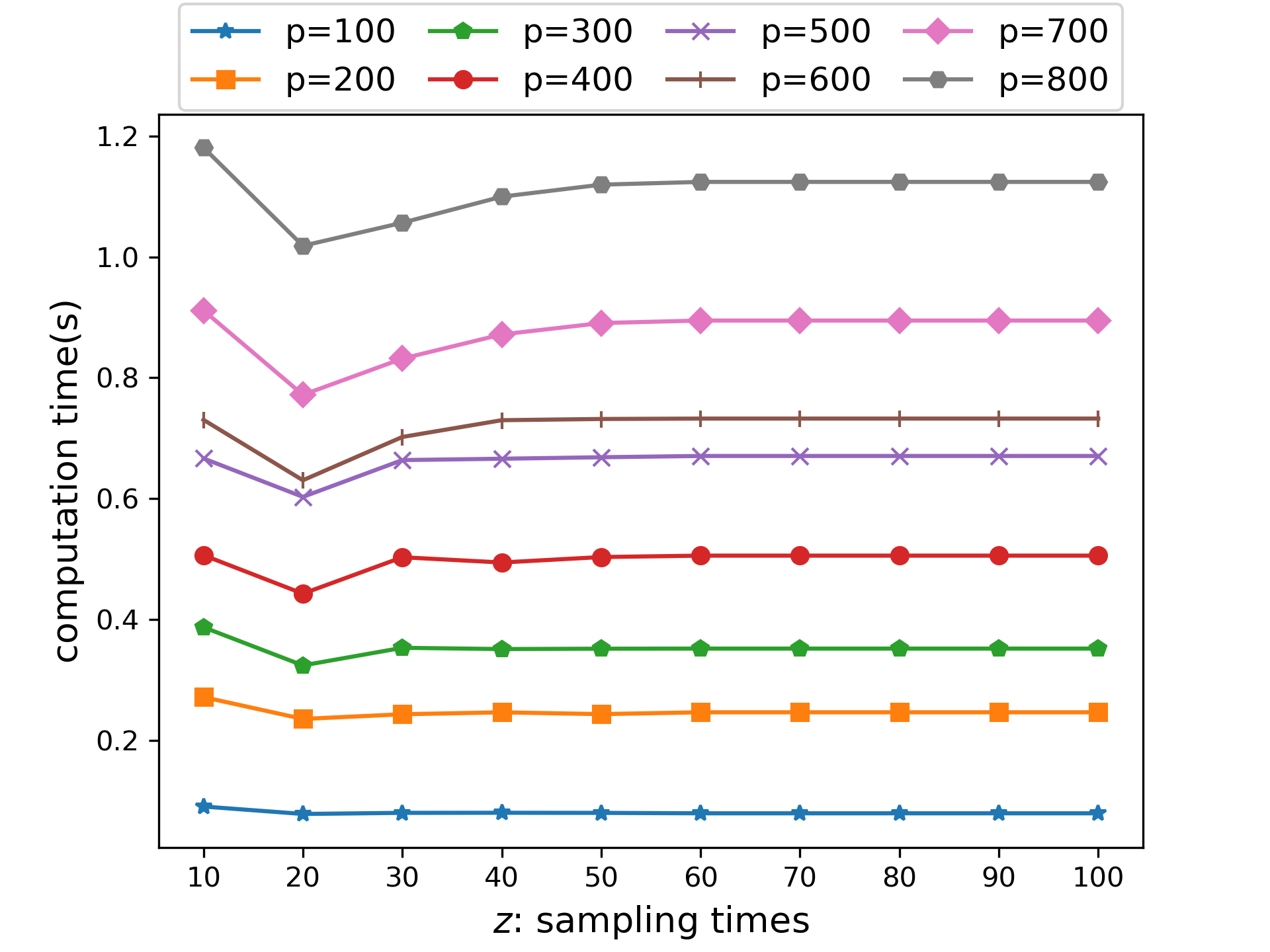}}
	\label{relation_time cost}
	\caption{Performance evaluation of \emph{DivCAR} w.r.t. $ (z,p) $. \label{relation_z_p}}
\end{figure*}
\subsection{Experimental Results}

In this subsection, we verify the superiority of our \emph{DivCAR} with respect to the following four profiles, where $ K $ is equal to 10 in all cases.

\vspace{3 pt}
\noindent \textbf{\emph{Profile-1: Performance convergence evaluation of DivCAR w.r.t. sampling times $ z $}}
\vspace{3 pt}

In our DivCAR, the number of sampling, i.e., $ z $ in the sampling process of \textbf{\emph{Step 1}}, serves a fairly significant role in making diversified, efficient and accurate recommendations. The more sampling times, the higher the probability of generating the precise recommendations for a set of required keywords. Nevertheless, the resulting consumption will also be high. Thus, to evaluate if and how much \emph{DivCAR} trades accuracy for efficiency, we first statistically analyze the convergence performance of representive metrics, i.e., precision and diversity, under the influence of parameter $ z $ to guide our follow-up experiments.

In examining Figure \ref{figconvergence}, sampling times $ z $ are varied from 10 to 100, and each curve represents a parameter pair $ (p, r) $, where $ p $ falls into $ \left\lbrace200, 300, 400, 500, 600, 700, 800\right\rbrace $ and $ r $ belongs to $ \left\lbrace2, 3, 4, 5, 6\right\rbrace $. Running data on precision and diversity are reported in Figure \ref{figconvergence} (a) and Figure \ref{figconvergence} (b), respectively. The MP and MILD performance are relatively slow and basically stable at about 0.3 and 0.9 when parameter $ z $ grows to 100, respectively. Therefore, our subsequent experiments are based on the fact that $ z $ is equal to 100 instead of no greater value of $ z $.

\vspace{3 pt}
\noindent \textbf{\emph{Profile-2: Diversity comparison of the five methods}}
\vspace{3 pt}

As we discussed in the previous section, diversity is the top priority of our research. On the one hand, a lower diversity may lead to loss of users' satisfaction degree and interests. On the other hand, a larger diversity may consequently increases its recommendation success rate. To measure the diversity performance of \emph{DivCAR} in terms of the interlist diversity (MILD) metric, we vary value of $ r $ from 3 to 6 in this test. Figure \ref{profile-MILD} presents the averaged statistical data of the exported results with $ z = 100 $, $ p \in \left\lbrace100, 200, 300, 400, 500, 600, 700, 800\right\rbrace $. Likewise, the following three profiles, i.e., profile-3, profile-4 and profile-5 are also averaged.

As indicated in Figure \ref{profile-MILD}, not surprisingly, the value of mean interlist diversity (MILD) of our \emph{DivCAR} is obviously superior to \emph{SSR} by 65.81$\%$, \emph{KC\_MulAGR} by 22.71$\%$ and \emph{ATD\_JSC} by 22.8$\%$ on average across four cases with different $ r $, respectively. This comes from the sampling mechanism introduced in our \emph{DivCAR}, which makes good use of \emph{DivCAR}'s ability to levarage the randomness of the sampling process to guarantee the nonrepeatability of web APIs across distinct recommendation lists. To a certain extent, this avoid the so-called local optimum dilemma in the optimization problem. In addition, we can see that the baseline \emph{SSR} remains largely lower than other three methods, which confirms that the diversity of its algorithm needs to be further improved.  In contrast with this, the MILD values of \emph{KC\_MulAGR} and \emph{ATD\_JSC} are lower than that of \emph{DivCAR} since \emph{KC\_MulAGR} only utilizes the minimum group Steiner tree algorithm without sampling technique. And thus, as for improving the dilemma of local optimum, they don't work very well. Furthermore, from Figure \ref{profile-MILD}, it can be drawn that the performance of our method is comparable to that of the \emph{MSD} method in terms of MILD value, but the accuracy of \emph{MSD} was significantly worse than that of our proposal.

Figure \ref{profile-MILD} also shows that, when the number of $ r $ rises, the overall MILD data of these five approaches roughly decline. For example, one of the most obvious is that \emph{KC\_MulAGR} decrease from 76.32$\%$ to 58.92$\%$ and \emph{ATD\_JSC} decrease from 75.98$\%$ to 58.91$\%$. This is mainly because of the fact that more web APIs increase the likelihood of repetition among them. Therefore, more distinct web APIs can be recommended to app developers to improve the satisfaction and serendipity of recommendations through our \emph{DivCAR}. \emph{DivCAR} offers significantly global diversity of the web APIs across individual recommendation lists by sampling to give equal opportunities to both popular and less popular web APIs.

\vspace{3 pt}
\noindent \textbf{\emph{Profile-3: Compatibility comparison of the two methods}}
\vspace{3 pt}

Compatibility, as another goal of our study we need to guarantee, affects how many success rate a developer executes. In this experiment, since compatibility between web APIs is not considered in other two competitive methods \emph{ATD\_JSC} and \emph{SSR}, we only compare the compatibility of trees answered by \emph{DivCAR} with \emph{KC\_MulAGR} in terms of the mean inner-list compatibility (MILC) metric. Moreover, a larger compatibility of a tree means better compatibility among web APIs from the tree. Here, $ r $ is set to an integer between 2 and 6. The experimental data are shown in Figure \ref{profile-MILC}.

As exported in Figure \ref{profile-MILC}, there is no distinct tendency in MILC of \emph{DivCAR} and \emph{KC\_MulAGR} with the growth of $ r $. Their values basically fluctuate around 4. The reason for this phenomenon is that one or more web APIs are needed to collectively meet functional requirements represented by distinct number of keywords. Under normal conditions, more web APIs often lead to larger MILC values. In addition, there is little difference in MILC value between \emph{DivCAR} and \emph{KC\_MulAGR}. For example, \emph{DivCAR}'s MILC outperform \emph{KC\_MulAGR} by 0.2, 1.59, 1.52 and 0.89 when $ r $ is equal to 3, 4, 5 and 6, respectively. This shows another advantage of sampling technique to avoid global optimization. Therefore, our \emph{DivCAR} can always achieve high-level web APIs compatibility.

\vspace{3 pt}
\noindent \textbf{\emph{Profile-4: Accuracy comparison of the five methods}}
\vspace{3 pt}

In this part, we evaluate and compare the recommendation accuracy of five methods by measuring the MP and MR metrics, which are recognized as the key measurements for evaluating the probability of ``False-positive'' and ``False-negative'', respectively. The statistical experiment results are indicated in Figure \ref{profile_precision_recall}.

As displayed in Figure \ref{profile_precision_recall} (a) and Figure \ref{profile_precision_recall} (b), the MP and MR values of the three methods, \emph{DivCAR}, \emph{KC\_MulAGR}, \emph{SSR} and \emph{MSD}, all increase with the growth of $ r $. For instance, \emph{DivCAR}'s MP and MR gradually increase by 9$\%$ from 23$\%$ to 32$\%$ in Figure \ref{profile_precision_recall} (a) and by 10$\%$ from 32$\%$ to 42$\%$ in Figure \ref{profile_precision_recall} (b), respectively. This is because the validity of recommended web APIs will increase with the rise of required web APIs to fulfill more complex functional requirements for an app. Here, what needs to be explained is that the MP and MR data of \emph{DivCAR} are superior to those of \emph{SSR} and \emph{MSD}, but perform slightly worse than \emph{KC\_MulAGR} and \emph{ATD\_JSC}. This comes from two main reasons. On the one hand, the fact our \emph{DivCAR} outperforms \emph{SSR} originates from that we utilize minimum group Steiner tree algorithm. On the one hand, there is a tradeoff between accuracy and diversity needs to be adjusted according to the needs of different scenes. To better demonstrate why such balance between the accuracy and diversity is the right balance, we defined a harmonic mean of the diversity and accuracy according to F2-score, which is calculated by $ (1+4)\frac{MP*MILD}{(4*MP)+MILD } $. The values of harmonic mean are 0.6107, 0.6082, 0.6002, 0.1672, 0.3826 for DivCAR, KC\_MulAGR, ATD\_JSC, SSR and \emph{MSD}, respectively. Especially, our focus is on the diversity of web API name in research scenarios of our paper. To be specific, except for \emph{SSR}, although the accuracy of \emph{KC\_MulAGR} and \emph{ATD\_JSC} is better than that of \emph{DivCAR}, their diversity is not as good as that of \emph{DivCAR}. For example, \emph{DivCAR} obtains significant merits over \emph{SSR}, i.e., 65.81$\%$, 18$\%$ and 22.13$\%$ in MILD, MP and MR, respectively; \emph{DivCAR} outperforms \emph{ATD\_JSC} in MILD by 22.8$\%$ on average, but is inferior to \emph{ATD\_JSC} in MP and MR by 21.6$\%$ and 22.62$\%$. It's also important to point out here that although the diversity of our \emph{DivCAR} is comparable to that of \emph{MSD}, the accuracy is significantly better than that of \emph{MSD}, i.e., 16$\%$ and 17$\%$ on average in MP and MR. Nonetheless, more importantly, the accuracy of \emph{DivCAR} is still able to meet the needs of developers in most cases. Therefore, the performance of our algorithm can still be guaranteed.

\vspace{3 pt}
\noindent \textbf{\emph{Profile-5: Efficiency comparison of the five methods}}
\vspace{3 pt}

Efficiency, as an important metric to evaluate algorithm performance, is tested and made a comparison of five different recommendation methods. The consumed time cost of five approaches is illustrated in Figure \ref{profile-timecost}.

As demonstrated in Figure \ref{profile-timecost}, the time consumption of the five methods all increase with the number of $ r $. Among them, the fastest growth is especially \emph{SSR} so that it grows almost linearly. This comes from the fact that more query keywords often require more complicated search processes. Moreover, the time cost of \emph{SSR} presents an approximately and linearly positive correlation with $ r $, since time consumption is generated when candidate web APIs are clustered into $ r $ distinct categories in the first stage of the algorithm. Among these approaches, the most time-consuming method is \emph{MSD} as it concerns the sum of the diversity of all combinations. With the growth of $ r $, both \emph{KC\_MulAGR} and \emph{DivCAR} methods consume more time to find the top-\emph{K} appropriate web APIs compositions from increasing numbers of candidate web APIs. \emph{KC\_MulAGR} needs more or less as much consumed time as \emph{ATD\_JSC} does as they all first generate all candidate result trees and then to find diverse top-\emph{K} web APIs recommendation lists. However, it is obvious from Figure \ref{profile-timecost} that the growth in consumption time of \emph{DivCAR} is slightly more significant than that of \emph{KC\_MulAGR} and \emph{ATD\_JSC}. This is because slightly more time is required to find the top-\emph{K} optimal solutions from fewer sampled nodes in the first step of our \emph{DivCAR}. Plus, the excellent diversity value of \emph{DivCAR}, the time cost is perfectly acceptable.

\vspace{3 pt}
\noindent \textbf{\emph{Profile-6: Recommendation performance evaluation of our DivCAR w.r.t. (z, p)}}
\vspace{3 pt}

In our algorithm, the two parameters, i.e., sampling times $ z $ and sampling size $ p $, can affect the exported recommendation results. To investigate their influence on \emph{DivCAR}'s performance, $ z \in \left\lbrace 10, 20, 30, 40, 50, 60, 70, 80, 90, 100 \right\rbrace $ and $ p \in \left\lbrace 100, 200, 300, 400, 500, 600, 700, 800 \right\rbrace $. Therefore, we test the average accuracy, diversity, compatibility and efficiency performance of our \emph{DivCAR} across one hundred cases with different $ z $ - $ p $ combinations in terms of MP, MILD, MILC and computation time, respectively. The experimental results are shown in Figure \ref{relation_z_p}, in which each line represents a change trend in the performance of $ p $ at different $ z $ and can all converges when $ z = 100 $.

As Figure \ref{relation_z_p} indicates, as $ z $ grows, the increase in $ z $ from 10 to 100 significantly impacts the values of MP obtained by different sample sizes. In Figure \ref{relation_z_p} (a), compared with $ z = 10 $, \emph{DivCAR} achieves much higher performance when $ z \textgreater 10 $. For instance, on average across different cases with  $ z \textgreater 10 $, \emph{DivCAR}'s MP outperforms $ z \textgreater 10 $ by 55$\%$. As presented Figure \ref{relation_z_p} (b), \emph{DivCAR} is also significantly affected along with the increase in $ z $, but in the opposite negative way. \emph{DivCAR}'s MILD slightly decreases by 3.34$\%$ on average. On one hand, it shows that the increases in MP can achieve much more significant than the such slight decreases in the values of MILD. On the other hand, we can conclude that the MP and MILD performances of our \emph{DivCAR} can reach the best case when the samping size $ p $ is 100. This mainly comes from that as the number of sampling times increases, more “appropriate web APIs” in each recommendation list inevitably result in less diversity of web APIs.

By contrast, the increase in $ z $ does not significantly impact \emph{DivCAR}’s compatability and efficiency measured by MILC and computation time, which further illustrates the stability of our method. More specifically, in the case of a small $ z $ at the beginning, the compatibility is not pretty stable, which is in line with our expected idea. But with the growth of $ z $, $ p $ exactly affects MILC and computation time. Under different $ p $, all the values of MILC vary from 2 to 22 in Figure \ref{relation_z_p} (c) and  the values of computation time all vary from 0.1 to 1.1 in Figure \ref{relation_z_p} (d). These changes are perfectly acceptable range. In addition, computation time gradually escalates with the growth of sampling size $ p $, which perfectly validates our expected results. This is because the search processes become more complicated as the number of nodes in sampled subgraphs increase. With $ p = 100 $, \emph{DivCAR} produces the most significant merit over other different $ p $, i.e., 0.089 seconds in computation time. Lastly, it is worth noting in particular that the finding mentioned in the classic paper \cite{Leskovec2006Sampling} that sampling strategy can achieve ideal effects with size down to approximately 15$\%$ of original massive graph, which exactly corresponds to the case $ z = 100 $ and $ p = 100 $ across all combinations.

\section{Conclusion}

Web APIs recommendation in IoT settings has become a promising way for app developers to develop desirable apps quickly and effciently. However, the results demonstrate that existing web APIs recommendation algorithms still suffer from low diversity. To overcome this issue, in this paper, we propose \emph{DivCAR} by means of the idea of game theory in IoT and sampling technique, to achieve diversity-aware and compatibility-driven web APIs recommendation for mashup development in IoT. In \emph{DivCAR}, we employ a random walk sampling technique on a ``API-API'' correlation graph prebuilt from ``APP-API'' co-usage records to generate diverse ``API-API'' correlation subgraphs. Afterwards, with the diverse ``API-API'' correlation subgraphs, we model the compatible web APIs recommendation problem as a minimum group Steiner  tree search problem; moreover, through solving the problem, manifold sets of compatible and diverse web APIs are made available to the app developers. At last, extensive experiments based on a real-world dataset from \emph{programmableWeb} validate the effectiveness and efficiency of our proposed \emph{DivCAR} approach.

In our future work, the quality data of IoT web APIs, i.e., response time, will be extended into our algorithm to imporve the recommendation accuracy. In addition, we will leverage more additional information of apps and web APIs from \emph{programmableWeb}, e.g., their descriptions and versions, for more practical and diverse apps in IoT.

%


\bibliographystyle{elsarticle-num}
\bibliography{reference}







\end{document}